\documentclass{egpubl}
 \JournalPaper         % uncomment for final version of Journal Paper
 \electronicVersion % can be used both for the printed and electronic version

\ifpdf \usepackage[pdftex]{graphicx} \pdfcompresslevel=9
\else \usepackage[dvips]{graphicx} \fi

\PrintedOrElectronic

\usepackage{t1enc,dfadobe}

\usepackage{egweblnk}
\usepackage{cite}

\newcommand{\bx}{{\mathbf{\bm x}}}
\newcommand{\bu}{{\mathbf{\bm u}}}

\usepackage{bm} % for bold greek symbols
\usepackage{amsmath}
\usepackage{overpic}
\usepackage{tikz}
\usetikzlibrary{backgrounds}

\title[Time-Dependent 2-D Vector Field Topology: An Approach Inspired by Lagrangian Coherent Structures]%
      {Time-Dependent 2-D Vector Field Topology: An Approach Inspired by Lagrangian Coherent Structures}

\author[F. Sadlo \& D. Weiskopf]
       {F. Sadlo and D. Weiskopf
 \thanks{\{sadlo,weiskopf\}@visus.uni-stuttgart.de}
        \\
	Visualization Research Center, Universit\"{a}t Stuttgart, Germany
       }

\begin{document}

\maketitle

\begin{abstract}
This paper presents an approach to a time-dependent variant of the concept
of vector field topology for 2D vector fields. Vector field topology is defined for steady vector
fields and aims at discriminating the domain of a vector field into regions
of qualitatively different behavior. The presented approach represents a
generalization
for saddle-type critical points and their separatrices to
unsteady vector fields based on generalized streak lines, with the classical vector field
topology as its special case for
steady vector fields.
The concept is closely related to that of Lagrangian coherent
structures obtained as ridges in the finite-time Lyapunov exponent field.
The proposed approach is evaluated on both 2D time-dependent synthetic
and vector fields from computational fluid dynamics. 

\begin{classification} % according to http://www.acm.org/class/1998/
\CCScat{I.3.8}{Computer Graphics}{Applications}
\CCScat{J.2}{Physical Sciences and Engineering}{Physics}

\end{classification}

\end{abstract}

%-------------------------------------------------------------------------
\section{Introduction and related work}

Vector field topology, introduced by Helman and Hesselink~\cite{rpbib:helman1989}, aims at the extraction of special stream lines related to singularities, most importantly the isolated zeros called \emph{critical points}.
Critical points, as well as 3D periodic orbits (closed stream lines), can be classified into attracting, repelling, or saddle-like types by analysis of the velocity gradient. Vector field topology is utilized in a wide field of applications, from fluid dynamics to the large field of continuous dynamical systems.
L\"offelmann et al.~visualize the behavior near the singularities of a 3D system
using various techniques such as glyphs
\cite{rpbib:loeffelmann1997streamarrows} and Poincar\'e maps \cite{rpbib:loeffelmann1998poincare}. 
The \emph{topological skeleton} is obtained by extracting singularities (critical points and periodic orbits), and if they are of saddle type, also the \emph{separatrices}, i.e.~their stable and unstable manifolds. Isolated periodic orbits can be extracted with the method of Wischgoll and Scheuermann~\cite{rpbib:wischgoll2001detection}.
In two dimensions the topological skeleton provides a segmentation of the domain into regions of qualitatively similar behavior.
In three dimensions, the 2D separating manifolds theoretically provide segmentation. However, in practical flows, these stream surfaces are often too convoluted~\cite{rpbib:peikert2007vortexrings} and lead to occlusion problems.
An alternative is to show only their pairwise intersections, 
known as saddle connectors~\cite{rpbib:theisel2004saddleconnectors}
or heteroclinic orbits,
showing the connectivity between critical points.
In three dimensions, there are two more types of singularities that can be investigated beyond what the topological skeleton provides, namely invariant tori~\cite{rpbib:peikert2009beyond} and strange attractors.
Another 
purpose of critical points, besides characterizing the neighborhood and computing skeletons,
is their use for seeding stream lines and stream surfaces, as has been done e.g.~by 
%L\"offelmann et al.~\cite{rpbib:loeffelmann1998dynamical} and 
Ye et al.~\cite{rpbib:ye05}.  
Even if 
critical points are 
used without any type analysis, this strategy was shown to yield effective visualizations by Weinkauf et al.~\cite{rpbib:weinkauf03a}.
Garth et al.~\cite{rpbib:garth2004surface} and Tricoche et al.~\cite{rpbib:tricoche2004intricate} demonstrated how complex flow structures such as vortex breakdown bubbles can effectively be visualized by using 
stream surfaces.

However, vector field topology
suffers from one important drawback: it only 
gives an instantaneous view on
vector fields because it is based on stream lines and is therefore not directly
interpretable for time-dependent vector fields.
The so-called Lagrangian coherent structures (LCS) represent
a time-dependent alternative.
%Haller~\cite{rpbib:haller2001distinguished,rpbib:haller2001lagrangian} 
Haller~\cite{rpbib:haller2001distinguished}
has shown that they can be formulated as ridges in
the 
field of the finite-time Lyapunov exponent \cite{rpbib:haller2001distinguished} (FTLE), measuring the divergence of trajectories, and Lekien et al.~\cite{rpbib:lekien2005} have
confirmed that LCS  
behave as
material lines (in 2D) or material surfaces (in 3D).
The FTLE-based approach has recently attracted much interest in the visualization community~\cite{rpbib:buerger2008,rpbib:soni2008},
however, it
cannot in an intrinsic way single out special points, classify them, and characterize their neighborhood.  There is however such a classification for material lines and surfaces, due to Haller~\cite{rpbib:haller2001distinguished}.
Unlike vector field topology, FTLE depends on the choice of the integration time and therefore,
resulting LCS are parameter-dependent.
A third, more practical limitation is that their computation is very costly, even if optimization, as proposed by Garth et al.~\cite{rpbib:garth2007} and Sadlo et al.~\cite{rpbib:sadlo2007efficient}, is used, because a trajectory has to be integrated for each sample. 
An ideal ``unsteady vector field topology'' would respect the Lagrangian view, but not have the limitations of the 
 FTLE-based approach.
This is of course an ambitious goal that will require solving several problems.
What we propose here is to look into the use of 
\emph{generalized streak lines} (GSL)~\cite{Wiebel07generalizedStreaks} 
as a replacement for
separatrices and critical points.
Experiments performed on analytic and CFD vector fields showed that GSL, if appropriately seeded, are often consistent with FTLE ridges.
More prominently, Haller \cite{Haller2000invariantManif} showed in 2000 that there exist so-called \emph{hyperbolic trajectories} with the property that neighboring path lines converge towards them in positive or negative time. We will reinterpret his findings in terms of vector field topology, one result being the fact that these hyperbolic trajectories are 
a
generalized counterpart to critical points for time-dependent vector fields and that GSL seeded at hyperbolic trajectories represent 
separatrices.
GSL have the additional advantage of being parameter-free (except for their geometric length) and easily computable by a method that generates lines directly, not via ridge extraction.

%-------------------------------------------------------------------------
\section{Classical vector field topology}
\label{sec:vft}

The concept of vector field topology for steady vector fields or
snapshots (single time steps) of time-dependent vector fields
 was originally defined by means
of special types of stream lines, or orbits in terms of dynamical systems theory:
\emph{critical points} and \emph{separatrices}.

Stream lines that degenerate to points because starting at zeros of the
vector field are called \emph{stationary points} and if they exhibit
a regular velocity gradient they are called critical points.
Critical points constitute the basis of vector field topology
and can be classified by means of the eigenvalues of the velocity
gradient at these locations into different categories. Mainly,
one discriminates \emph{saddle}-type critical points exhibiting
real eigenvalues of opposite sign (or negative determinant), 	
\emph{nodes} where signs are equal,	
and \emph{foci} with a pair of complex eigenvalues.	
In the special case of 2D divergence-free vector fields,
\emph{centers} are
structurally stable critical points, too.

Stream lines that converge to 
saddles
in either positive
or negative time form the separatrices which separate regions of
different behavior of the vector field. In 3D, a saddle has a pair of
a one-dimensional and a two-dimensional separatrix, i.e.~a stream line and a stream surface. The separatrices can be further classified into
\emph{stable} and \emph{unstable} manifolds, converging to the critical point in  positive or negative direction of time, respectively.
In 3D, additional separatrices are obtained from stream lines that converge to saddle-type periodic orbits in positive or negative time. They form a pair of 2D manifolds, a stable and an unstable one.

In the following section we motivate a time-dependent alternative for the two concepts of critical points and separatrices, and reinterpret the concept of hyperbolic trajectories and the trajectories that converge to them in these terms.

%-------------------------------------------------------------------------
\section{Generalized vector field topology}
\label{sec:gvft}

For steady-state vector fields, LCS defined by FTLE ridges coincide in many cases with manifolds in the sense of vector field topology \cite{rpbib:haller2001distinguished,rpbib:sadlo2009comparison}, stable manifolds corresponding to repelling LCS and unstable ones to attracting LCS \cite{Haller2000invariantManif}. Manifolds are attracting in the sense that a perturbation perpendicular to the manifold grows exponentially in reverse time direction and repelling in the sense that such a perturbation grows exponentially in forward time. FTLE usually exhibits ridges along separatrices because separatrices are converging to saddle-type critical points in the respective direction of time whereas nearby streamlines pass the critical point and hence typically end up at locations distant from the critical point, and hence lead to an increased FTLE.
Because LCS are material lines, this motivates to interpret separatrices as streak lines converging to respective saddle-type critical points in the appropriate direction of time.
As in the case of classical vector field topology, these separatrices can be constructed by seeding them at small offset from the respective critical point and developing them in the appropriate direction of time.
This leads to our approach to a time-dependent (generalized) vector field topology (GVFT): 
simply replace the role of stream lines by generalized streak lines \cite{Wiebel07generalizedStreaks} in the concept.
Since streak lines and stream lines are identical in steady vector fields, this generalization does not change the classical vector field topology but may provoke new interpretations.

In the case of time-dependent vector fields, on the other hand, GVFT and vector field topology will usually differ substantially. In contrast to classical separatrices, our generalized separatrices (the generalized streak lines) advect with the flow and our generalized critical points (degenerate generalized streak lines) may move over time. The fact that our generalized critical points may move over time necessitates to use generalized streak lines, a variant of streak lines where the seed is allowed to move.

The above reasoning about the correspondence between FTLE ridges and generalized separatrices (generalized streak lines) also holds for time-dependent vector fields: (moving) hyperbolic regions
% (loci of diverging trajectories) 
take over the seeding role of saddle-type critical points in the case of time-dependent vector fields. For the same reasons, the generalized streak line converges to that region in the respective direction of time even if the region is moving.
A major drawback with classical vector field topology is its frame dependence, i.e.~a Galilean transformation can arbitrarily move critical points and even lead to bifurcations, i.e.~the creation or destruction of critical points.
In the original approach \cite{rpbib:helman1989}, it is therefore mentioned that an appropriate frame of reference has to be chosen when applying the concept of vector field topology, surely a unsatisfactory circumstance in many cases.
Fortunately, FTLE and hence the concept of LCS are Galilean invariant and since we aim at an approach inspired by LCS, we also aim at Galilean invariance.
%Haller TODO:cite manif2000 defines r
Regions can be defined to be hyperbolic if the determinant of the velocity gradient has negative sign, hence a Galilean invariant property, and consistent with the classification of classical saddle-type critical points. However, on the one hand these regions are usually quite large, and on the other they can move at arbitrary speed in time-dependent vector fields leading to no substantial separation effect on neighboring trajectories passing the region. Additionally, we propose the generalized critical points to be degenerate generalized streak lines. Generalized streak lines are only degenerate if they are seeded on path lines in space-time, i.e.~the seed itself is a particle that advects with the flow.
All in all, this motivates us to define generalized critical points in the space-time view as distinguished path lines inside hyperbolic regions.

The concept of hyperbolic trajectories was already proposed in 2D by Haller \cite{Haller2000invariantManif} in 2000
% for 2D vector fields 
using the so-called hyperbolicity time: the time a trajectory spends in a hyperbolic region until it leaves it for the first time. 
He states that a trajectory is hyperbolic if two 1D manifolds of trajectories converge towards it, one in forward and one in reverse time direction.
For finding the seed for such a hyperbolic trajectory at a given time $t_0$, he proposed to compute the hyperbolicity time in forward and backward time direction, to detect the local maxima (which we interpret here as height ridges according to Eberly \cite{rpbib:eberly1994}) in these two scalar fields and to determine the intersections of these ridges.
If the ridges are ``sharp'', i.e.~thin and the deformation rate of the LCS low, these intersections already represent the seeds for 
hyperbolic trajectories, otherwise one intersects the ``flat'' ridge-like regions and obtains regions representing sets of candidates for hyperbolic trajectories, which need then to get further restricted by additional conditions in his theorem~1.

 Ide et al.~\cite{Ide2002distinguishedHyperbolicTraj} proposed
an alternative approach: they obtain candidates for their so-called distinguished hyperbolic trajectories from a temporal analysis of classical critical points and then find the distinguished hyperbolic trajectories by construction of a time-dependent linear model.
We restrict this work to Haller's formulation
and refer the reader to the dynamical systems and fluid mechanics literature for methods and discussions of how to find hyperbolic trajectories in general.
Haller addressed 2D time-dependent vector fields that are not necessarily divergence-free, interpreted time as third dimension, and stated that the trajectories converging to the hyperbolic trajectory form 2D manifolds in space-time (see Figure~\ref{fig:gyreSaddle-sineTransl-manifold-lines}). This is perfectly consistent with our approach because if these manifolds are intersected by a plane of constant time, one obtains exactly the generalized streak lines seeded on the hyperbolic trajectory in the appropriate direction of time.

The presented approach may be incomplete in the sense that no time-dependent counterpart in terms of
generalized streak lines is given for the classical critical points of type node, focus, and center, as well as for periodic orbits. The authors are not aware of it in the field of dynamical systems
% theory
and fluid mechanics, and because it would probably go beyond the scope of this paper, it is addressed as future work.

In the following Section~\ref{sec:method}, we present methods for obtaining the seeds for hyperbolic trajectories
and the construction of the space-time streak manifolds. Then, in Section~\ref{sec:results}, we validate our method and show results using different synthetic and CFD examples.

%------------------------------------------------------------------------- 
\section{Method}
\label{sec:method}

In this section we present methods for
% sampling hyperbolicity time and detecting the intersections of sharp ridges therein
obtaining seeds for hyperbolic trajectories
 (Section~\ref{sec:hyp-time-and-ridges}) and for generating the space-time streak manifolds (Section~\ref{sec:streak-manifolds}).

%- - - - - - - - - - - - - - - - - - - - - - - - - - - - - - - - - - - - -
\subsection{Extraction of seeds for hyperbolic trajectories}
\label{sec:hyp-time-and-ridges}

As stated by Haller \cite{Haller2000invariantManif}, hyperbolicity time has nice properties compared to FTLE. Whereas FTLE builds on a gradient between the end points of trajectories and is therefore highly sampling-dependent, hyperbolicity time can be evaluated per trajectory and a value at a sample point does therefore not depend on neighboring values. Second, it
is less dependent
on the advection time $T$ that is used for its evaluation, compared to FTLE. He states that increasing $T$ renders sharper and sharper results. Therefore, $T$ can be chosen such that it covers the complete available temporal domain of the data.
Haller proposed
to sample hyperbolicity time on a grid of initial conditions, i.e.~starting points for the trajectories. 
Hyperbolicity time $d_T$ of the vector field $\bu(\bx,t)$ at time $t_0$ and location $\bx_0$ computes as
\begin{align}
\label{eq:hyp-time}
d_T(\bx_0,t_0) = & \nonumber \\ 
\max\limits_{t \in [t_0, t_0+T]} & \{t| \det \nabla \bu(\bx(\tau;\bx_0),\tau)<0, \,\,\, t_0 \leq \tau < t\}
\end{align}
with $\bx(\cdot;\bx_0)$ being the trajectory started at $\bx_0$,
% and $t_0$,
in other words, the time a trajectory spends in a hyperbolic region until it exits it the first time. Haller also proposes to apply early termination of the integration of the trajectories, meaning that integration can be stopped as soon as the trajectory enters a non-hyperbolic region. This accelerates the computation typically substantially since vector fields usually exhibit a high amount of (moving) non-hyperbolic regions.

\begin{figure}[tb]
  \centering
\begin{overpic}[width=.48\linewidth]{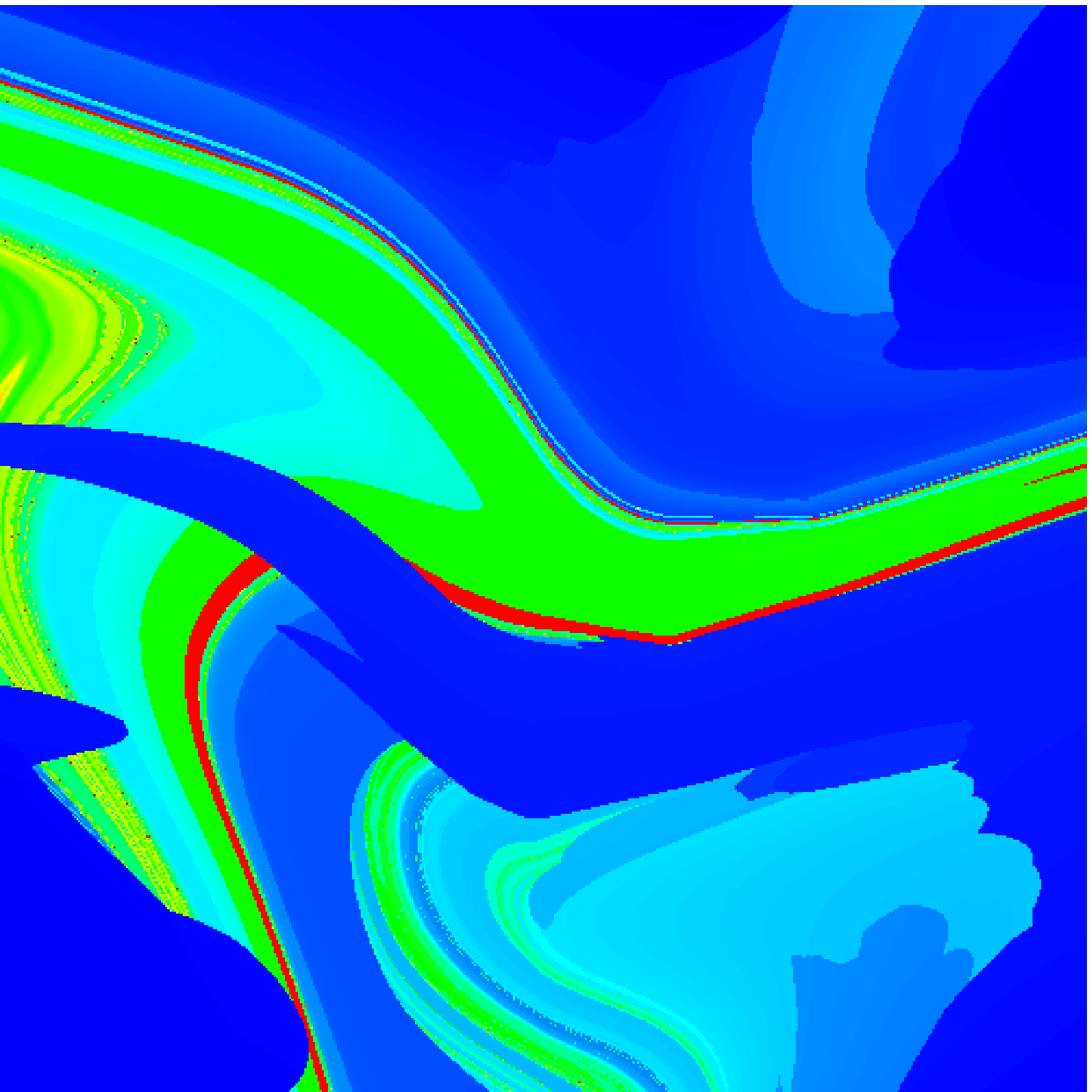}
\put(-4.4,8.5) {
\setlength\fboxsep{0pt}
\setlength\fboxrule{0.5pt}
\fbox{\includegraphics[width=0.06\linewidth]{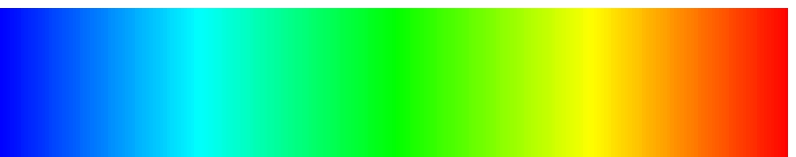}}
}
\put(0, 2.5) { \scriptsize{\textcolor{white}{$0$}}}
\put(11, 2.5) { \scriptsize{\textcolor{white}{$7$}}}
\put(14,70) {
  \begin{tikzpicture}
  \coordinate (A) at (.3,.1);
  \coordinate (B) at (0,0);
  \draw [->, white, very thick] (A) -- (B);
  \end{tikzpicture}
}
\put(41,29) {
  \begin{tikzpicture}
  \coordinate (A) at (.2236,.2236);
  \coordinate (B) at (0,0);
  \draw [->, white, very thick] (A) -- (B);
  \end{tikzpicture}
}
\end{overpic}
\begin{overpic}[width=.48\linewidth]{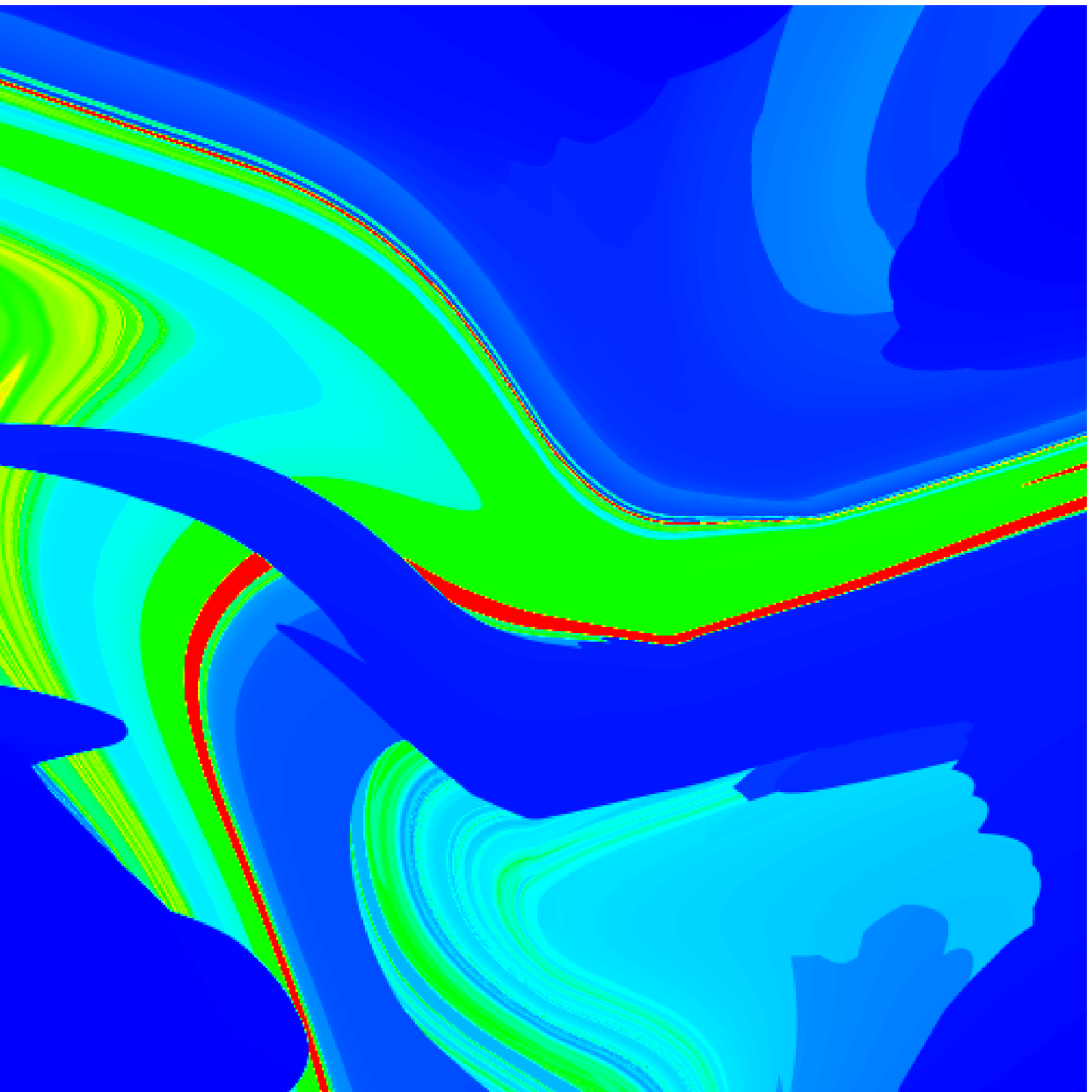}
\put(14,70) {
  \begin{tikzpicture}
  \coordinate (A) at (.3,.1);
  \coordinate (B) at (0,0);
  \draw [->, white, very thick] (A) -- (B);
  \end{tikzpicture}
}
\put(41,29) {
  \begin{tikzpicture}
  \coordinate (A) at (.2236,.2236);
  \coordinate (B) at (0,0);
  \draw [->, white, very thick] (A) -- (B);
  \end{tikzpicture}
}
\end{overpic}\\
\vspace{0.8mm}
\begin{overpic}[width=.48\linewidth]{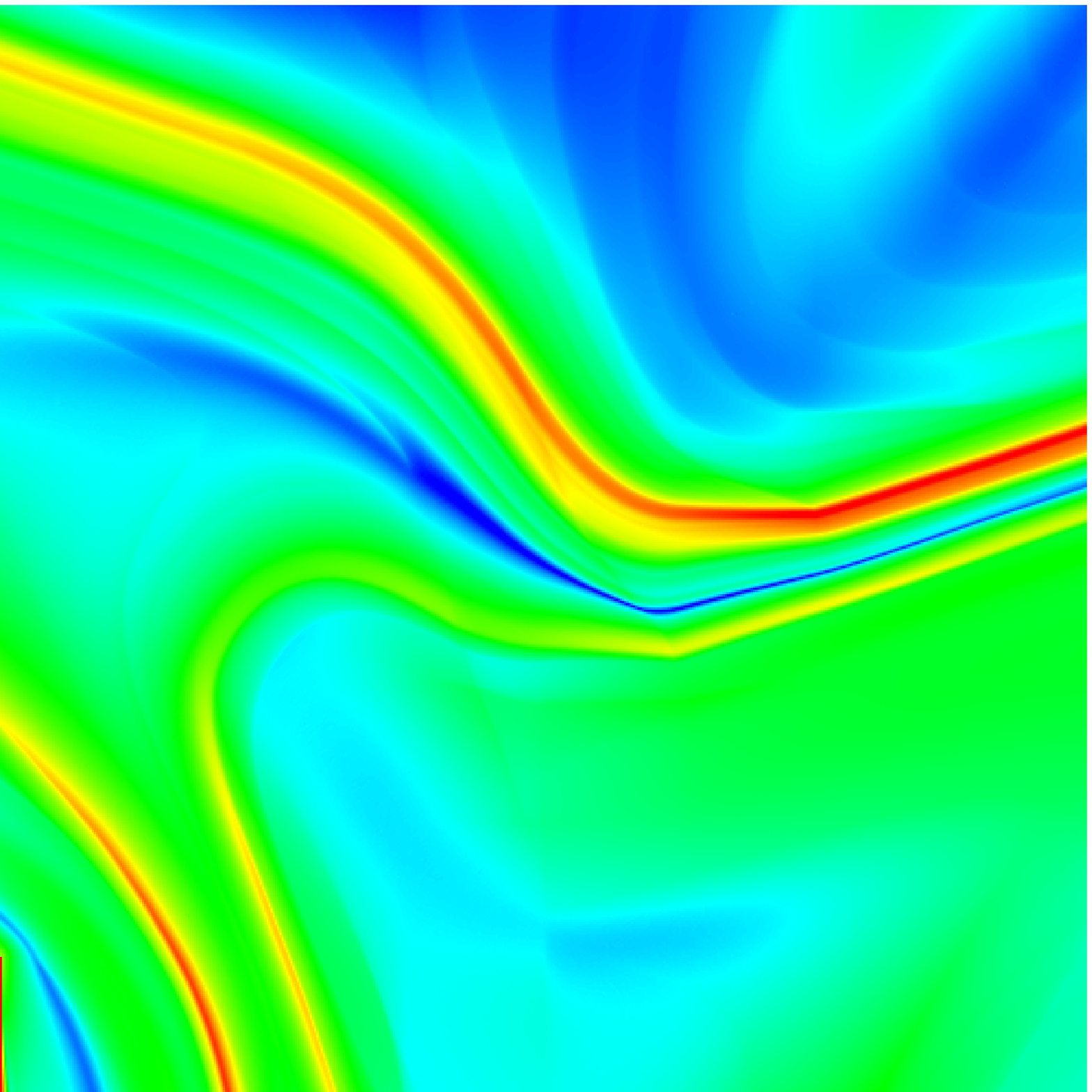}
\put(-4.4,8.5) {
\setlength\fboxsep{0pt}
\setlength\fboxrule{0.5pt}
\fbox{\includegraphics[width=0.06\linewidth]{figures/rainbow_cropped}}
}
\put(0, 2.5) { \scriptsize{\textcolor{black}{$0$}}}
\put(8.5, 2.5) { \scriptsize{\textcolor{black}{$11$}}}
\end{overpic}
\begin{overpic}[width=.48\linewidth]{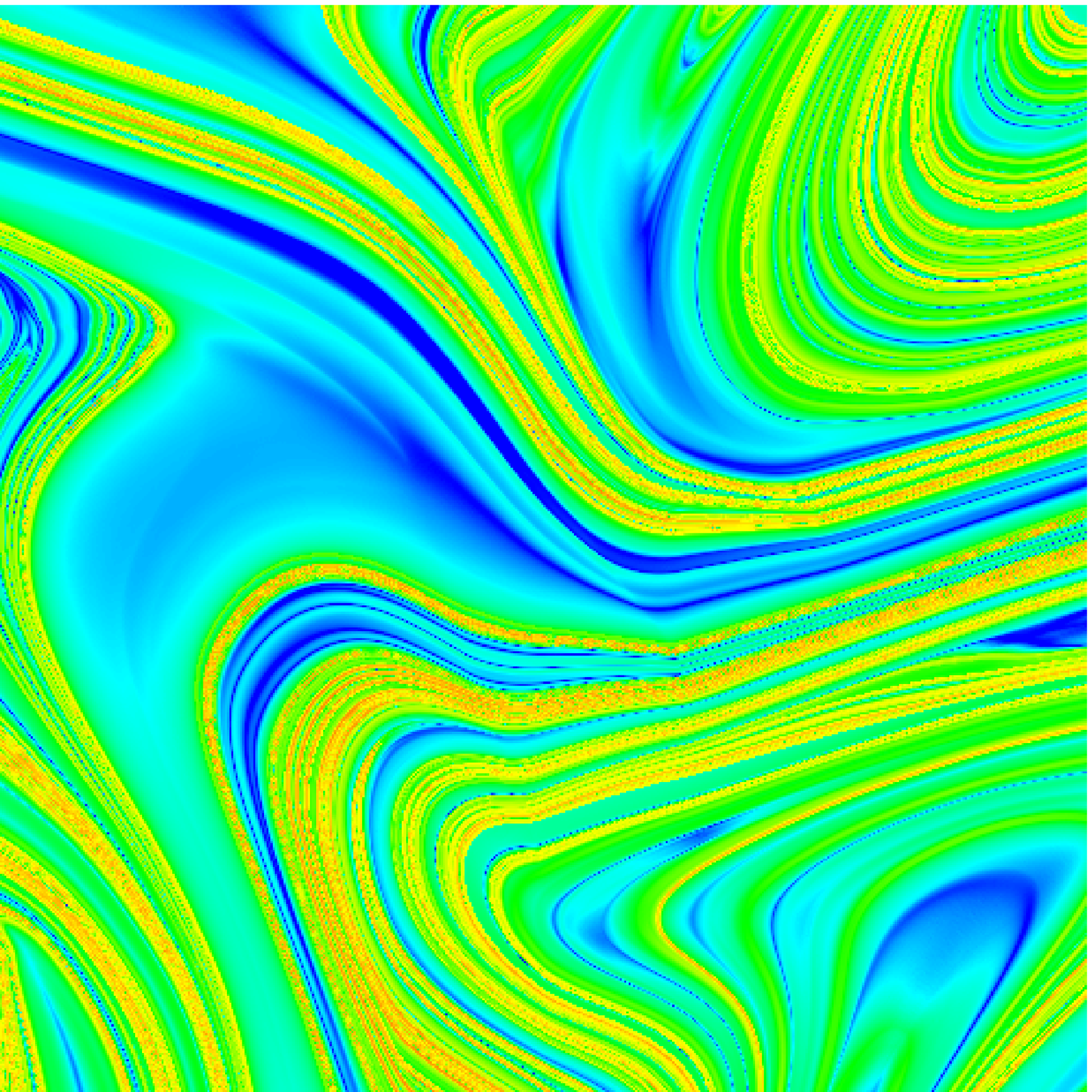}
\put(-4.4,8.5) {
\setlength\fboxsep{0pt}
\setlength\fboxrule{0.5pt}
\fbox{\includegraphics[width=0.06\linewidth]{figures/rainbow_cropped}}
}
\put(0, 2.5) { \scriptsize{\textcolor{black}{$0$}}}
\put(11, 2.5) { \scriptsize{\textcolor{black}{$3$}}}
\end{overpic}
  \caption{\label{fig:hyp-time-quantization}
           Forward hyperbolicity time in buoyancy dataset (see Section~\ref{sec:buoyancy}) inside time interval $[2, 9]$ computed without supersampling (top left, visible aliasing artifacts (arrows) at thin ridges) and with supersampling (top right). For comparison, FTLE at $t_0=2$ with advection time $T=0.5$ (bottom left) and $T=3$ (bottom right). }
\end{figure}

Figure~\ref{fig:hyp-time-quantization} (top row) shows a result for the buoyancy dataset subject to investigation in Section~\ref{sec:buoyancy}. There are mainly two issues with the sampling of hyperbolicity time. The first one is that one needs to evolve time (and hence the trajectory) in discrete steps in numerical data and a direct application of Eq.~\ref{eq:hyp-time}
in this context would lead to
quantization
by these time steps. Therefore, we detect the time step where it leaves the hyperbolic regime, or the domain, and perform bisection search
inside that time interval
 to determine
the exit time.
The second issue is aliasing. The ridges in hyperbolicity time are typically
extremely
thin and exhibit high gradients, therefore trajectories seeded on a regular grid often miss the features, leading to aliasing as shown in Figure~\ref{fig:hyp-time-quantization} (top left).
Increasing resolution does typically not reduce the problem
because the ridges are that thin.
 We perform supersampling to address this problem.
 Figure~\ref{fig:hyp-time-quantization} (top right) shows the same region of the field but this time $5 \times 5$ trajectories are seeded on a regular grid covering a pixel and the resulting hyperbolicity times are \emph{averaged}. This not only tends to produce contiguous and smooth ridges, it also results in reduced numerical noise. This is especially important since the extraction of height ridges is sensitive to noise \cite{rpbib:peikert2008ridge} because it relies on second derivatives.

The straightforward approach would be now to extract hyperbolicity time in forward and reverse time over a sufficiently large time interval to ensure sufficiently sharp ridges, extract ridges therefrom, and use the ridge intersections directly as seeds for hyperbolic trajectories, or, depending on the data, to additionally check if
the trajectories seeded at
 these intersections fulfill
over the interval $[t_0, t_0+T]$
 all conditions in Haller's theorem~1 \cite{Haller2000invariantManif}:
\begin{equation*}
\label{eq:theorem-1}
\det \nabla \bu(\bx(t),t)<0, \quad \sqrt{2}\beta \left[\frac{1}{\lambda_{1 \min}} + \frac{1}{\lambda_{2 \min}}\right] < \alpha
\end{equation*}
with
the real eigenvalues $- \lambda_1(t) < 0 < \lambda_2(t)$ of $\nabla \bu(\bx(t),t)$,
\begin{equation*}
 \lambda_{k \min} = \min\limits_{[t_0, t_0+T]} \lambda_k(t), \quad k=1,2
,
\end{equation*}
 $M$ being the matrix of the temporally consistently oriented eigenvectors of $\nabla \bu(\bx(t),t)$,  and
\begin{equation*}
 \alpha = \min\limits_{t \in [t_0, t_0+T]} |\det M|, \quad
\beta = \max\limits_{t \in [t_0, t_0+T]} \left|\frac{\partial}{\partial t} M \right|.
\end{equation*}
 With
\begin{equation*}
\label{eq:gamma}
\gamma = \frac{\sqrt{2} \beta [ \alpha^{2} \lambda_{1 \min} \lambda_{2 \min} + \sqrt{2} \alpha \beta ( \lambda_{1 \min} + \lambda_{2 \min}) + 2 \beta^{2}]} {\alpha^{3} \lambda_{1 \min} \lambda_{2 \min}}
,
\end{equation*}
the additional conditions of Haller's theorem~1 for a trajectory being \emph{uniformly finite-time hyperbolic} inside the interval $[t_0, t_0+T]$ are:
\begin{align*}
\beta & < \frac{\alpha}{2} \sqrt{2 \lambda_{1 \min} \lambda_{2 \min}},\\
\lambda_1(t) & > \gamma + \frac{2 \beta^2}{ \alpha^2 \lambda_{1 \min} \lambda_{2 \min}} \lambda_2(t),\\
\lambda_2(t) & > \gamma + \frac{2 \beta^2}{ \alpha^2 \lambda_{1 \min} \lambda_{2 \min}} \lambda_1(t).
\end{align*}

Although intersecting ridges of hyperbolicity time works well for some cases such as our oscillating gyre-saddle example (Section~\ref{sec:gyre-saddle-oscillating}), we encountered problems even for the simple quad-gyre example (Section~\ref{sec:quad-gyre}), and in particular in case of CFD
% simulations 
%results
(Section~\ref{sec:buoyancy}). 
Several problems can arise: 
\begin{itemize}
\item
The temporal domain of the field may not allow hyperbolicity time to develop sharp enough ridges, meaning that regions of constant maximum time (compare red band in Figure~\ref{fig:hyp-time-quantization} (top row)) remain, which do not allow for the robust extraction of height ridges.
In this case, after intersection of forward and backward hyperbolicity time candidate sets, a subset of these adjacent samples represents candidates for uniformly hyperbolic trajectories and many of them may fulfill the additional conditions of theorem~1, resulting in adjacent uniformly hyperbolic trajectories, instead isolated ones.
Extracting ridges would have the advantage that the ridge intersections are not restricted to the sampling grid, because ridge extraction is based on interpolation. Hence, ridge intersections would result in a much better approximation for the seeds, even if the ridges are very thin and hence would be hard to sample on a grid.
\item
The field may lead to intricate ridges, as in the case of the quad-gyre example (see below), making the extraction and intersection of ridges intractable.
\item
The ridge intersections may not be consistent with FTLE ridges for a desired FTLE advection time. This is basically no problem but the goal of this work is to derive a concept that is consistent with LCS in terms of FTLE ridges. 
\end{itemize}
Since both, ridges in the FTLE (of long enough advection time) and ridges in hyperbolicity time represent invariant manifolds, we prefer to extract ridges from FTLE fields and intersect those. As already mentioned, FTLE depends on the advection time used for its computation. This first seems as a drawback, but it also offers a scale-dependent approach: if the advection time is increased, FTLE ridges (the LCS) typically also get sharper but at the same time they often get longer and more convoluted (see Figure\ref{fig:hyp-time-quantization} (bottom row)). Intersecting these massively folded LCS leads to a huge number of intersections and may lead to insignificant visualizations. Hence, the advection time for FTLE computation can serve as a scale parameter. This motivation is supported by the observation that we were not able to extract uniformly hyperbolic trajectories, using both ridges in hyperbolicity time and FTLE, in CFD data in significant regions and time intervals, as discussed in Section~\ref{sec:buoyancy}.
Additionally, rejecting FTLE ridge regions if the FTLE falls below a threshold allows to quantitatively restrict the analysis to LCS of a required strength of
repulsion or attraction.
Furthermore, FTLE is, due to its spatially variational definition, typically smooth even at low resolutions and does not exhibit flat regions in non-degenerate vector fields, hence its ridges are well defined and precise with respect to the chosen time scale. Further, we
% usually
typically
impose a threshold on the intersection angle between ridges to suppress intersections of almost parallel ridges that would lead to inaccurate and also often insignificant results.

If the data allow for it (depending on the rate at which LCS deform \cite{Haller2000invariantManif}, the available temporal domain, and numerics), we restrict the intersections to seeds for uniformly hyperbolic trajectories by applying the additional conditions of Haller's theorem~1. Otherwise, we propose
a
 ``weak'' hyperbolicity approach: we generate the space-time streak manifolds as long as the seeding trajectory is inside a hyperbolic region and stop their generation if it enters a non-hyperbolic region (or start the reverse-time manifold only at this time). This can be seen as a complementary approach to filtering the trajectories by their hyperbolicity time. Although these trajectories typically deviate from the intersection curve of the attracting and repelling invariant manifolds, this is not so much of a problem because the generated streaks get quickly attracted by the attracting manifold in the respective direction of time (compare Figure~\ref{fig:gyreSaddle-sineTransl3}
and
Figure~\ref{fig:buoyancy-1}~(bottom)).

Accurate seeds can be extracted by filtered AMR ridge extraction \cite{rpbib:sadlo2007efficient}, however at the risk of missed ridges and hence intersections. In our tests, we iteratively decrease the size of a sampling window of constant resolution around each intersection of the ridges until the change of the intersection falls below a tolerance.
To constrain the computational cost to an acceptable level
and to obtain significant visualizations
 in case of massively folded ridges, we supervised the search in these cases by adapting the ridge filtering criteria.
 We address an advanced algorithm for the seed search in case of massively folded ridges as future work, possibly based on a histogram analysis of the underlying field, also because our tests indicate that even highly accurate seeds with respect to the underlying FTLE or hyperbolicity time field do not
guarantee exact hyperbolic trajectories
over long time intervals
 (Figure~\ref{fig:gyreSaddle-sineTransl3} (bottom)), in the sense that they follow the intersection curves of the invariant attracting and repelling manifolds. 
It was already stated by Haller \cite{Haller2000invariantManif} that integrating hyperbolic trajectories is a hard task because there is a repelling manifold in each direction of time and therefore errors tend to grow exponentially.

\begin{figure}[tb]
  \centering
\begin{overpic}[width=.48\linewidth]{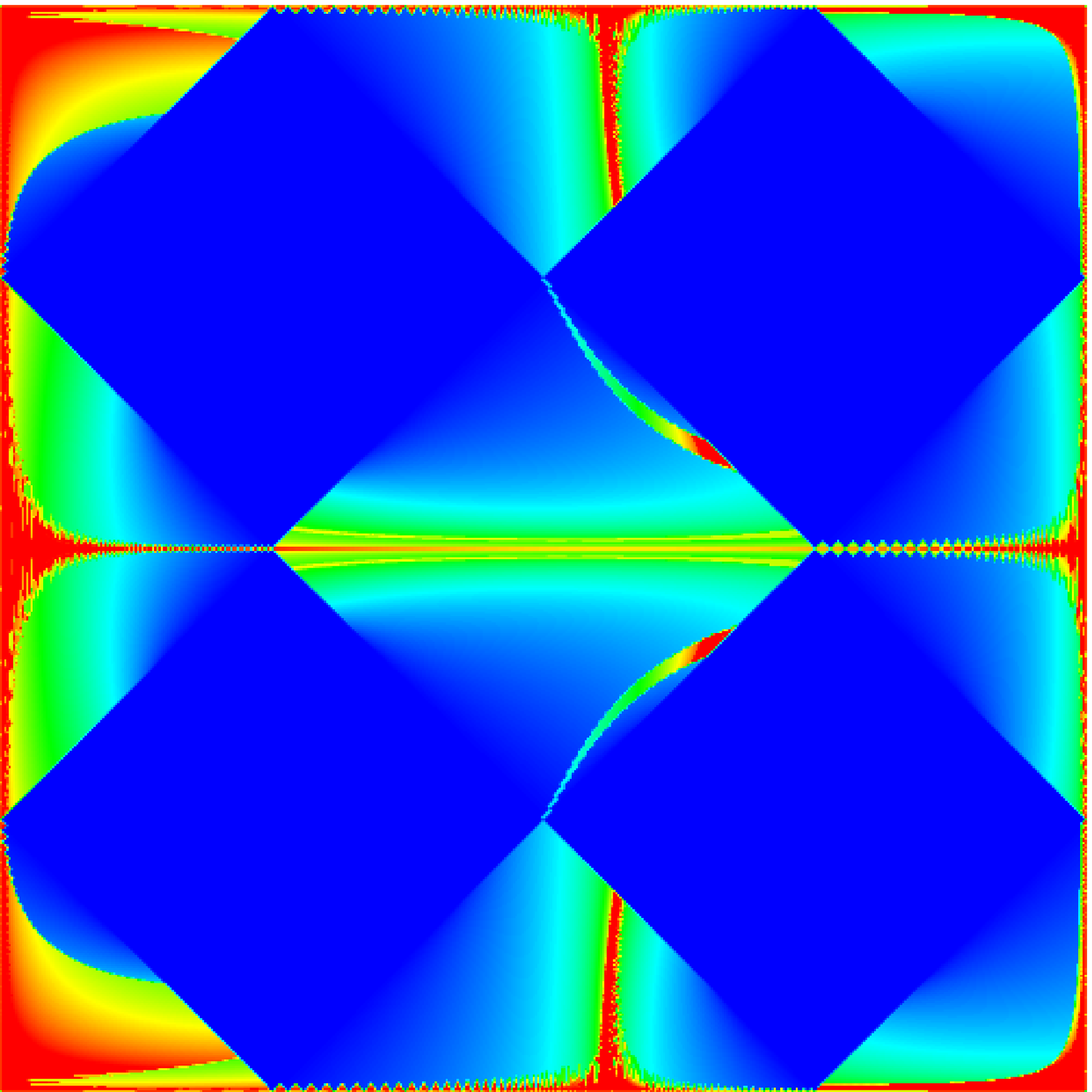}
\put(-4.4,8.5) {
\setlength\fboxsep{0pt}
\setlength\fboxrule{0.5pt}
\fbox{\includegraphics[width=0.06\linewidth]{figures/rainbow_cropped}}
}
\put(0, 2.5) { \scriptsize{\textcolor{black}{$0$}}}
\put(11, 2.5) { \scriptsize{\textcolor{black}{$7$}}}
\end{overpic}
\begin{overpic}[width=.48\linewidth]{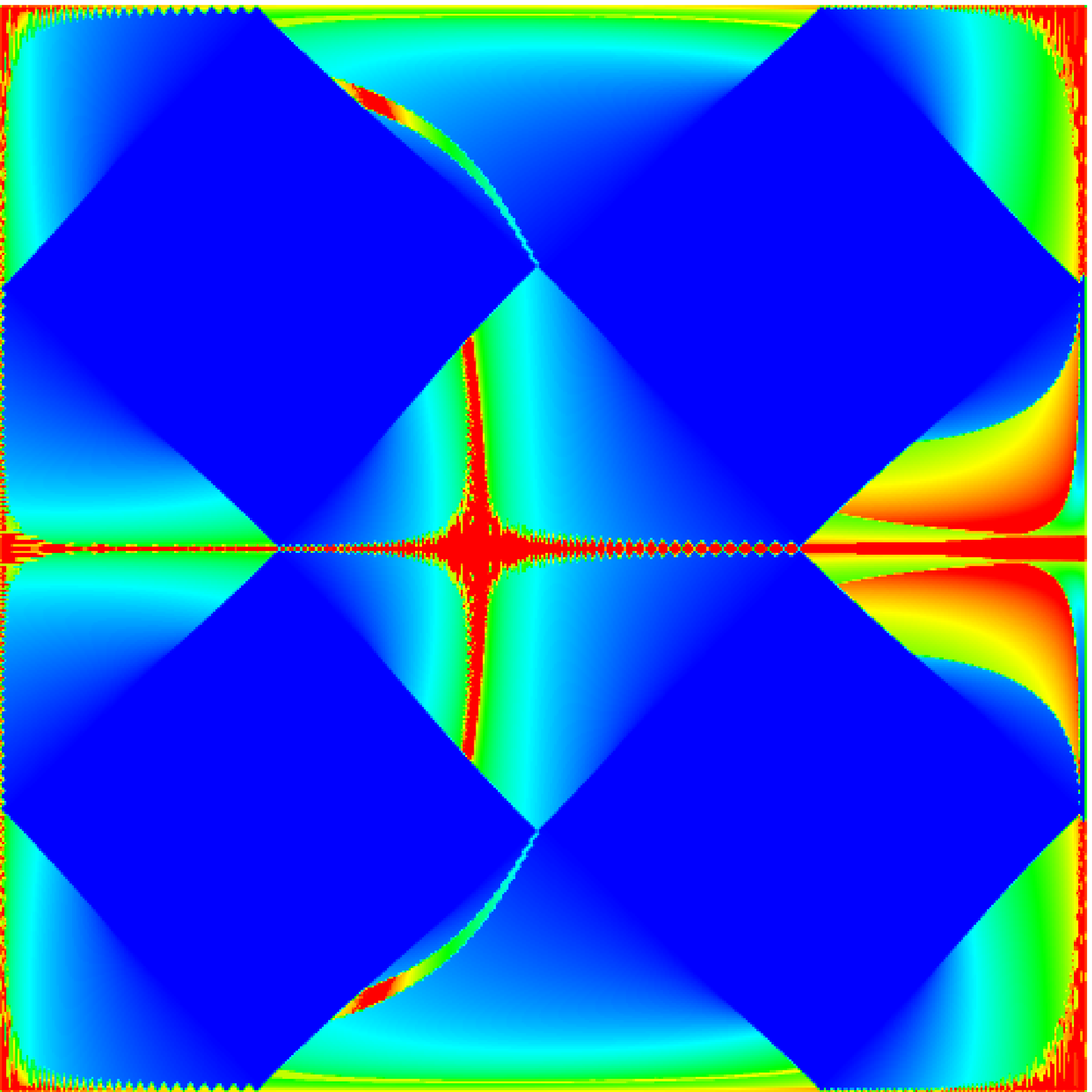}
\put(45,53) {
  \begin{tikzpicture}
  \coordinate (A) at (.2236,.2236);
  \coordinate (B) at (0,0);
  \draw [->, white, very thick] (A) -- (B);
  \end{tikzpicture}
}
\end{overpic}\\
\vspace{0.44mm}
\includegraphics[width=.48\linewidth]{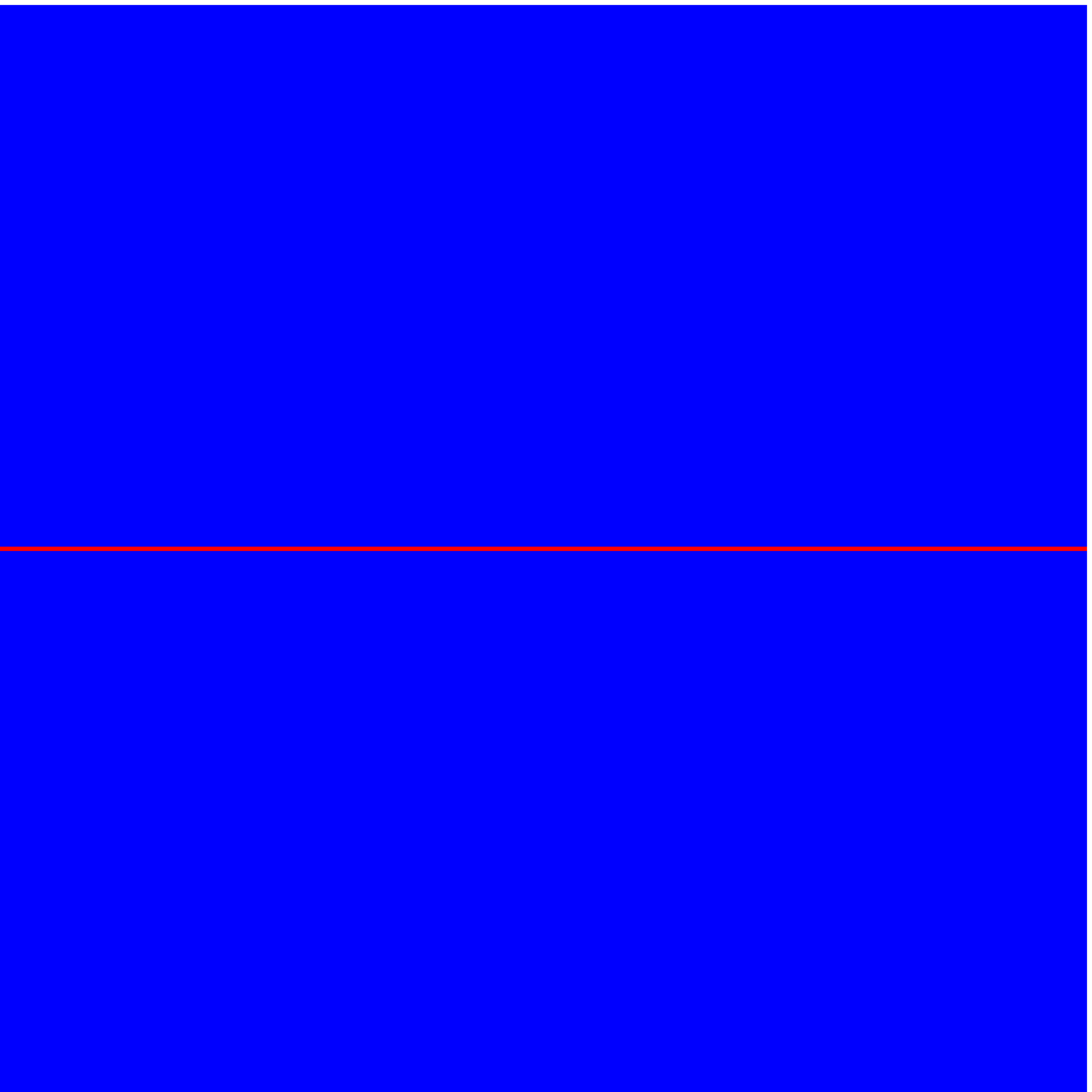}
\includegraphics[width=.48\linewidth]{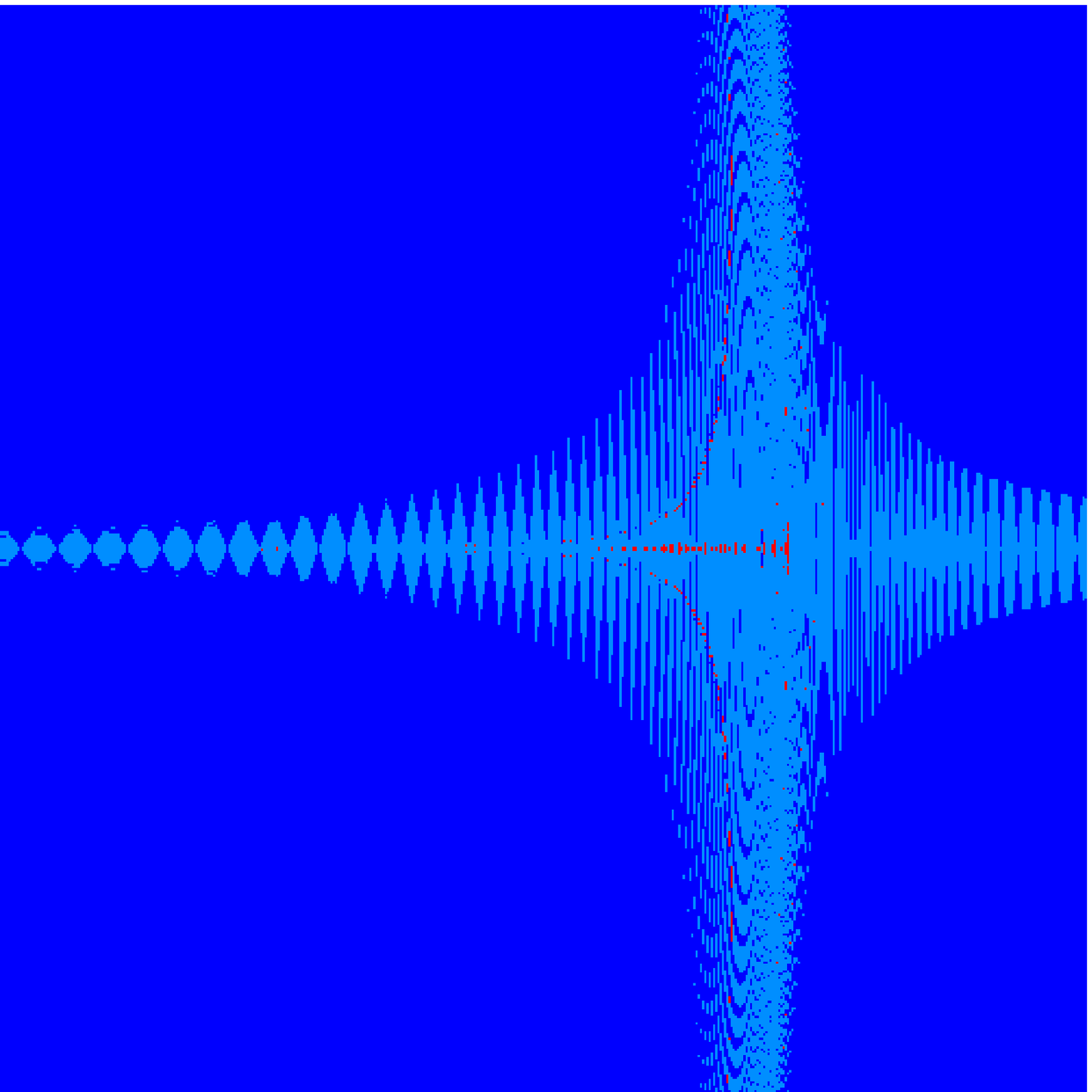}\\
\vspace{0.44mm}
\begin{overpic}[width=.48\linewidth]{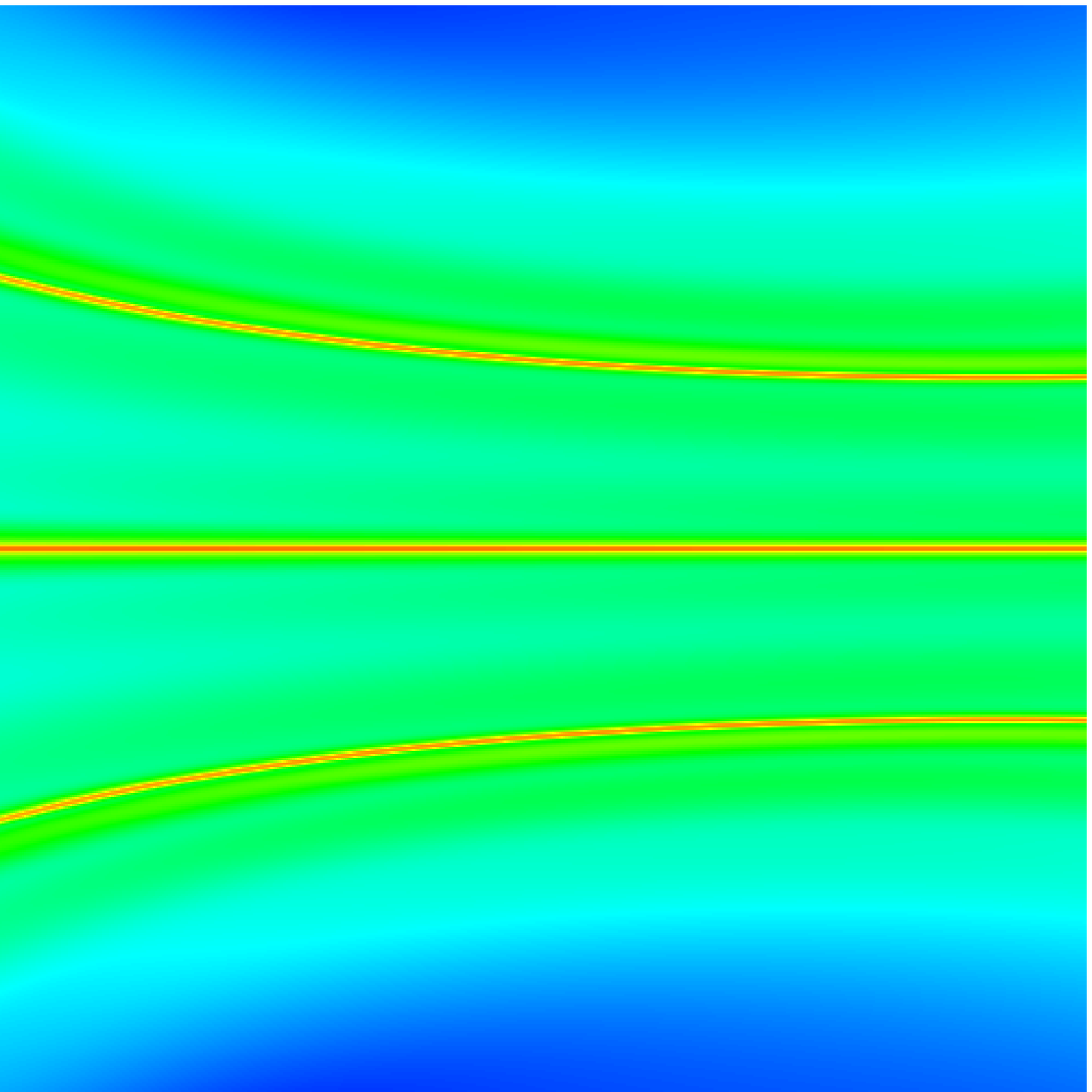}
\put(-4.4,8.5) {
\setlength\fboxsep{0pt}
\setlength\fboxrule{0.5pt}
\fbox{\includegraphics[width=0.06\linewidth]{figures/rainbow_cropped}}
}
\put(0, 2.5) { \scriptsize{\textcolor{black}{$0$}}}
\put(6.5, 2.5) { \scriptsize{\textcolor{black}{$0.8$}}}
\end{overpic}
\includegraphics[width=.48\linewidth]{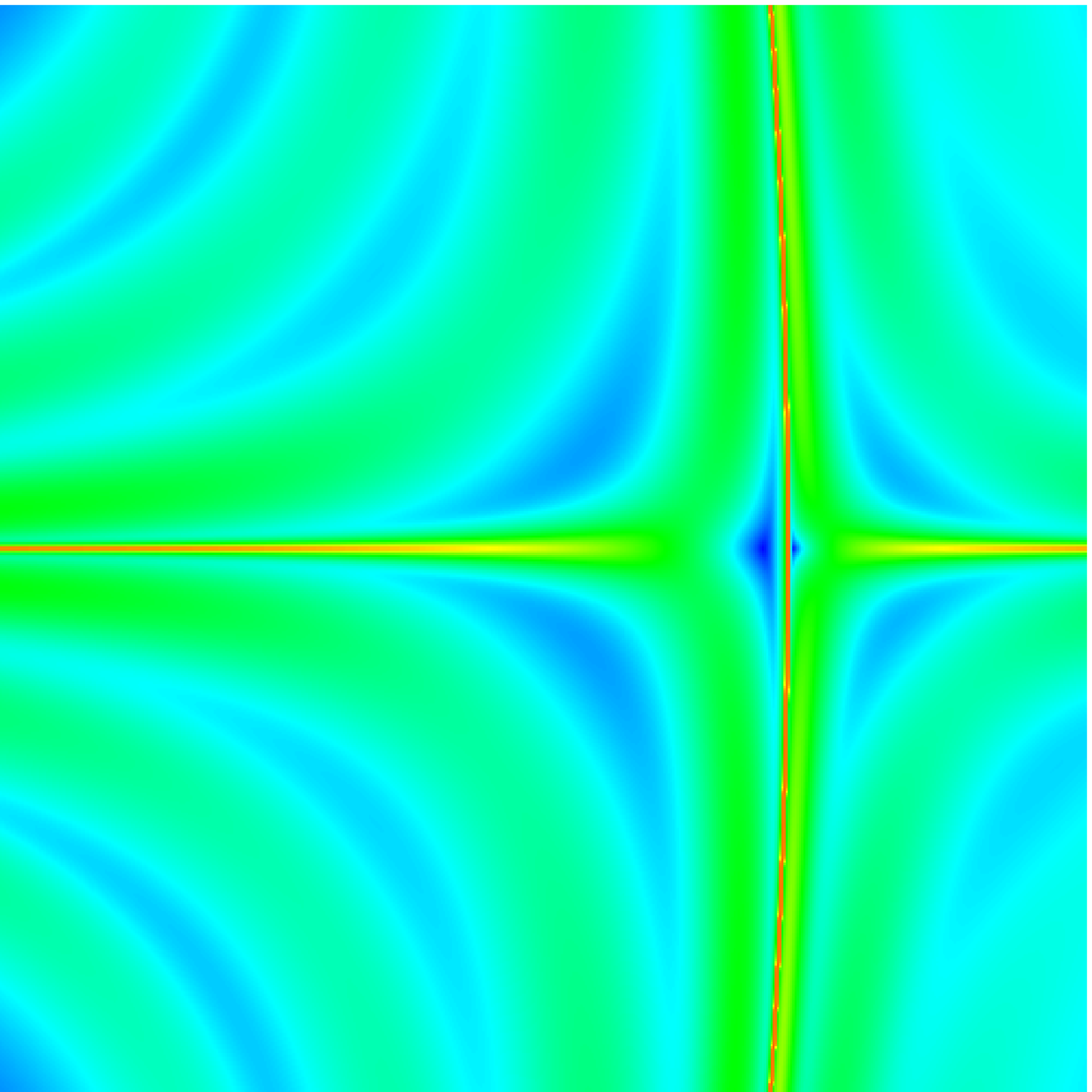}
  \caption{\label{fig:quad-gyre-hyp-time}
           Seed extraction in quad-gyre example at time $t_0=20$. Forward hyperbolicity time (top left) and backward (top right) of the complete region. Close-up region of uniform hyperbolicity (red) for time interval $[20,30]$ (mid left) and $[10,20]$ (mid right) with max-time hyperbolicity (light blue) around the
% central 
``cross-shaped'' high backward hyperbolicity time region
% (top right), 
(arrow)
and corresponding FTLE of advection time $T=10$ (bottom left) and $T=-10$ (bottom right). The backward uniform hyperbolicity
% image 
exhibits a small vertical line consistent with the LCS in backward-time FTLE and indeed, the intersection of  the FTLE ridges results in an accurate seed for a uniformly hyperbolic trajectory (Figure~\ref{fig:quad-gyre-1}). }
\end{figure}

To conclude this section, we present the problems and the chosen approach for extracting the seed for the quad-gyre example (Section~\ref{sec:quad-gyre}). Figure~\ref{fig:quad-gyre-hyp-time} (top row) shows its hyperbolicity time forward, and in reverse time direction. Whereas the forward hyperbolicity time behaves well in the central region where the seed has to be located, backward hyperbolicity time exhibits ``resonances'': there is a very high number of vertical ridges in the region where the aimed ridge is located (Figure~\ref{fig:quad-gyre-hyp-time} (top right)). This phenomenon forbids to extract ridges in this region and zooming as well as supersampling did not lead to a successful extraction of the desired (vertical) ridge. Therefore we also visualized inside a region around the presumed ridge intersection the points of uniform hyperbolicity according to all conditions of Haller's theorem~1 (Figure~\ref{fig:quad-gyre-hyp-time} (mid row)). Whereas the forward hyperbolicity time exhibits a single horizontal ridge, backward hyperbolicity still exhibits a multitude of seeds for uniformly hyperbolic trajectories  along that line. Therefore we have chosen to extract the ridges from FTLE (Figure~\ref{fig:quad-gyre-hyp-time} (bottom row)), which resulted in a seed for an accurate uniformly hyperbolic trajectory.

%- - - - - - - - - - - - - - - - - - - - - - - - - - - - - - - - - - - - -
\subsection{Extraction of space-time streak manifolds}
\label{sec:streak-manifolds}

\begin{figure}[tb]
  \centering
\begin{overpic}[width=1\linewidth]{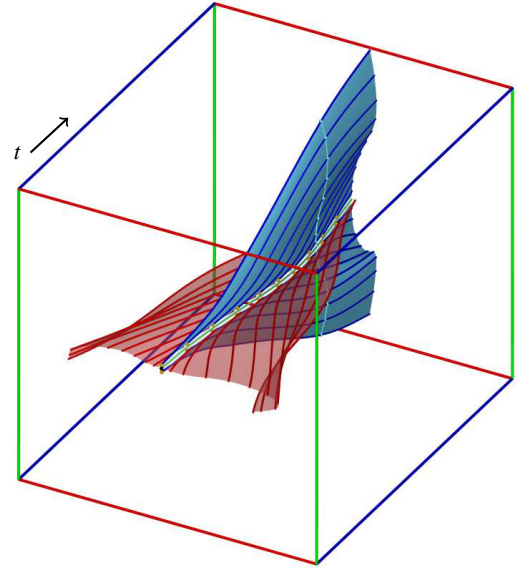}
\put(3,70.5) {
  \begin{tikzpicture}
  \coordinate [label=left:$t$] (A) at (0,0);
  \coordinate (B) at (0.5,0.4625);
  \draw [->, thick] (A) -- (B);
  \end{tikzpicture}
}
\end{overpic}
  \caption{\label{fig:gyreSaddle-sineTransl-manifold-lines}
           Attracting invariant manifold (light blue) and repelling invariant manifold (red) consist of trajectories (blue, red) converging to the hyperbolic trajectory (green tube seeded at black sphere) in the respective direction of time. Space-time streak manifold generation:
% Hyperbolic trajectory (green tube) seeded at black sphere. 
attracting streak manifold
% (light blue) 
is seeded in forward direction along hyperbolic trajectory with offset in eigenvector direction of the velocity gradient at the respective position and time (orange spheres). The space-time streak manifold consists of the trajectories (blue) of each particle of the two generalized streak lines. An isotemporal slice of the manifold represents the two generalized streak lines (turquoise).
}
\end{figure}

As explained in Section~\ref{sec:gvft}, the generalized streak lines representing time-dependent separatrices, or manifolds, are seeded along hyperbolic trajectories. In a space-time view (regarding time as third dimension of the 2D vector field), the generalized streak lines become 2D manifolds (e.g.~Figure~\ref{fig:gyreSaddle-sineTransl-manifold-lines}) which we call space-time streak manifolds. 

Several techniques were proposed for the extraction of the invariant stable and unstable manifolds
% (generalized streak lines in our sense) 
of hyperbolic trajectories.
Because the 
%streak lines 
manifolds
usually undergo massive thinning (stretching) and also possibly folding,
% due to the manifolds of the hyperbolic trajectory, 
refinement of the resulting 
%streak 
lines
% (and hence the resulting space-time streak surfaces)
 is usually necessary.
 Mancho et al.~\cite{Mancho2003manifolds} give an overview of different existing approaches and also present new ones. In general, they initially seed particles along the eigenvector of the velocity gradient (the eigenvector corresponding to the chosen direction of time), discuss different criteria for deciding when a new trajectory (in our sense a new streak particle) needs to get inserted during advection for obtaining manifolds 
without ``gaps'', and insert
required particles at the previous time step by interpolation.

% We do not follow the Hultquist family of intergal surface construction algorithms (inserting seeds at the front by interpolation) but for better results, we seed by linear space-time interpolation along the seeding trajectory (the generalized critical point).
%
We follow a different approach: for a given direction of time, we seed two generalized streak lines at an offset from the hyperbolic trajectory (Figure~\ref{fig:gyreSaddle-sineTransl-manifold-lines}). A seed for each streak line is generated at regular time steps along the hyperbolic trajectory and the offset is orientated along the eigenvector of the velocity gradient at the respective position and time of the hyperbolic trajectory,
the largest eigenvalue for the extraction of attracting manifolds (integration in forward time) and the smallest for repelling manifolds (integrated in reverse time).
%Hereby, 
We use a user-defined constant offset distance from the hyperbolic trajectory.
This offset distance has to be small enough to allow for the linearization by the velocity gradient, also with respect to temporal change. However, because the manifold is attracting in the respective direction of time, errors tend to shrink
and therefore tend to be negligible. It is also necessary to orientate the eigenvector to be consistent with the respective streak in order to prevent ``switching'' of the seeds.

At each time step of the space-time streak manifold generation procedure, we measure the distance between
neighboring
particles of each streak line and test it against a user-defined threshold. If this 
threshold is violated and hence insertion of a particle is required,
we do not interpolate between already advected streak particles,
similar to the Hultquist family of integral surface construction algorithms \cite{Hultquist1992surface,rpbib:garth2004surface,garth2008surfaces}, 
 but linearly interpolate between the corresponding seed positions in space and time and use the interpolated seed for generating a new trajectory up to the current time.
% The positions along the new trajectory are inserted into the space-time streak manifold. 
Additionally, to avoid excessive computation, we allow to erase streak particles if the distance falls below an other threshold. Hereby, we either allow only the deletion of particles that were generated by interpolation of the seed (and hence less accurate), or also allow the deletion of particles generated by ``true'' seeds. 
 
The space-time streak manifolds are terminated as soon as the corresponding hyperbolic trajectory enters a non-hyperbolic region. From that point on the remaining part of the trajectory may be visualized to support the spatial perception of where the space-time streak manifold ends (see Figure~\ref{fig:buoyancy-1} (bottom)). The corresponding reverse-time manifold is constructed along the same hyperbolic trajectory with the difference, that if the forward-time trajectory was stopped because it entered a non-hyperbolic region, the reverse-time manifold is constructed only from this time on (see Figure~\ref{fig:buoyancy-1} (bottom)). Of course, the manifolds can be constructed from hyperbolic trajectories integrated in both directions of time starting at the seeds. However, in this work we only show the manifolds for the hyperbolic trajectories computed in forward direction from the seeds.

%------------------------------------------------------------------------- 
\section{Results}
\label{sec:results}

In the following sections, we examine our approach and compare it to the approach by Haller \cite{Haller2000invariantManif} using different 2D vector fields, both synthetic and from CFD. The seeds for the hyperbolic trajectories were all obtained by ridge intersections of FTLE, not hyperbolicity time.

%- - - - - - - - - - - - - - - - - - - - - - - - - - - - - - - - - - - - -
\subsection{Skewing gyre-saddle example}
\label{sec:crit-non-moving}

One of the simplest cases in the field of LCS is a saddle-type critical
point that stands still, exhibits 
approximately constant eigenvalues and eigenvectors over time,
and persists over the
entire temporal domain.
Although this example may sound academic, it is quite common in many areas.
The probably most prominent and most important area are dynamical systems where it is quite common that the phase space exhibits non-moving critical points.
As another example, 
fluid dynamics solutions at low Reynolds numbers tend to converge to
quasi-steady state
%flow due to viscous effects and dissipation,
flow,
 provided
that there are constant boundary conditions.
% and no oscillating phenomena such as those provoking von K\'{a}rm\'{a}n vortex streets.

\begin{figure}[htb]
  \centering
  \includegraphics[width=.49\linewidth]{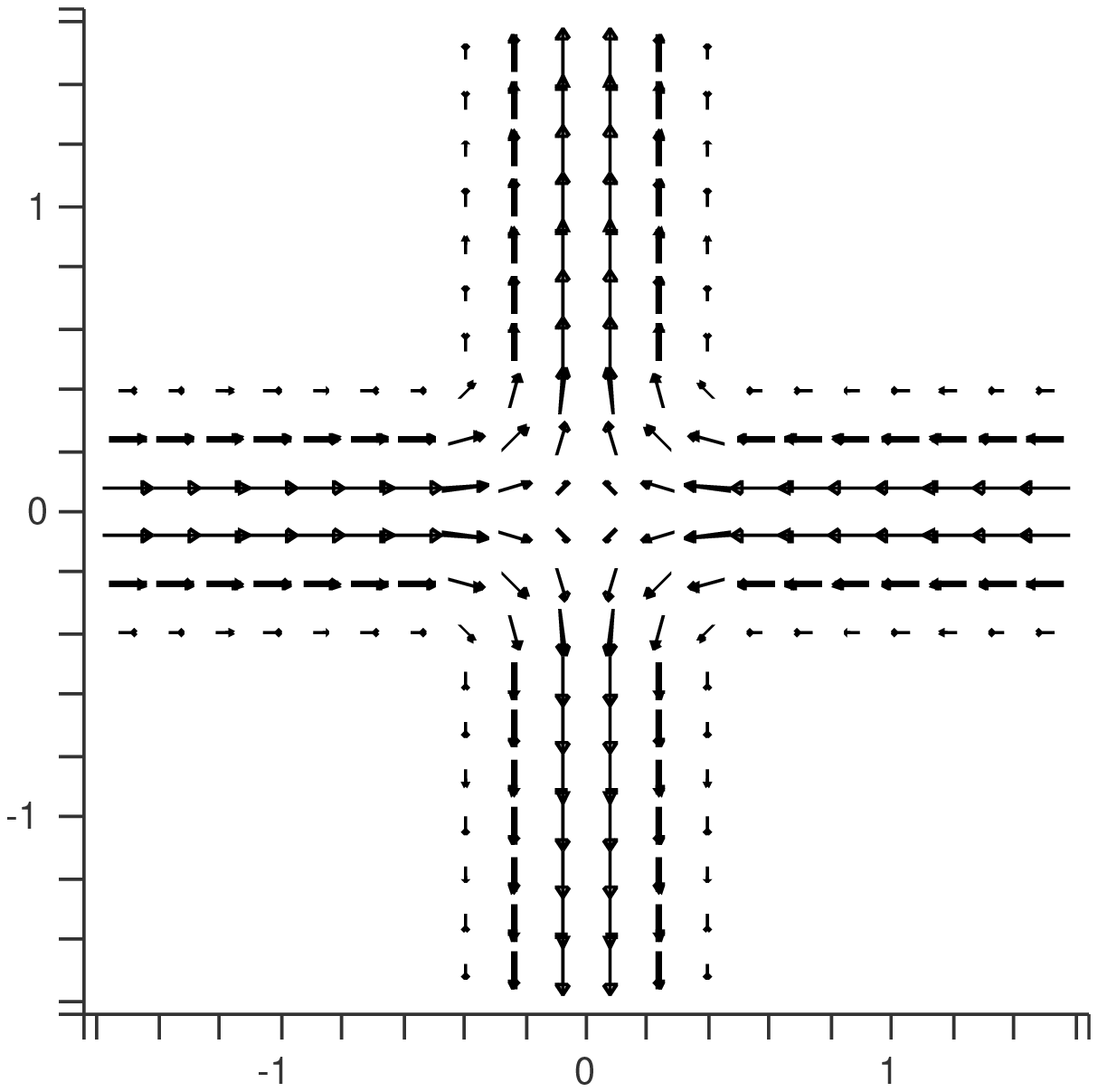}
  \includegraphics[width=.49\linewidth]{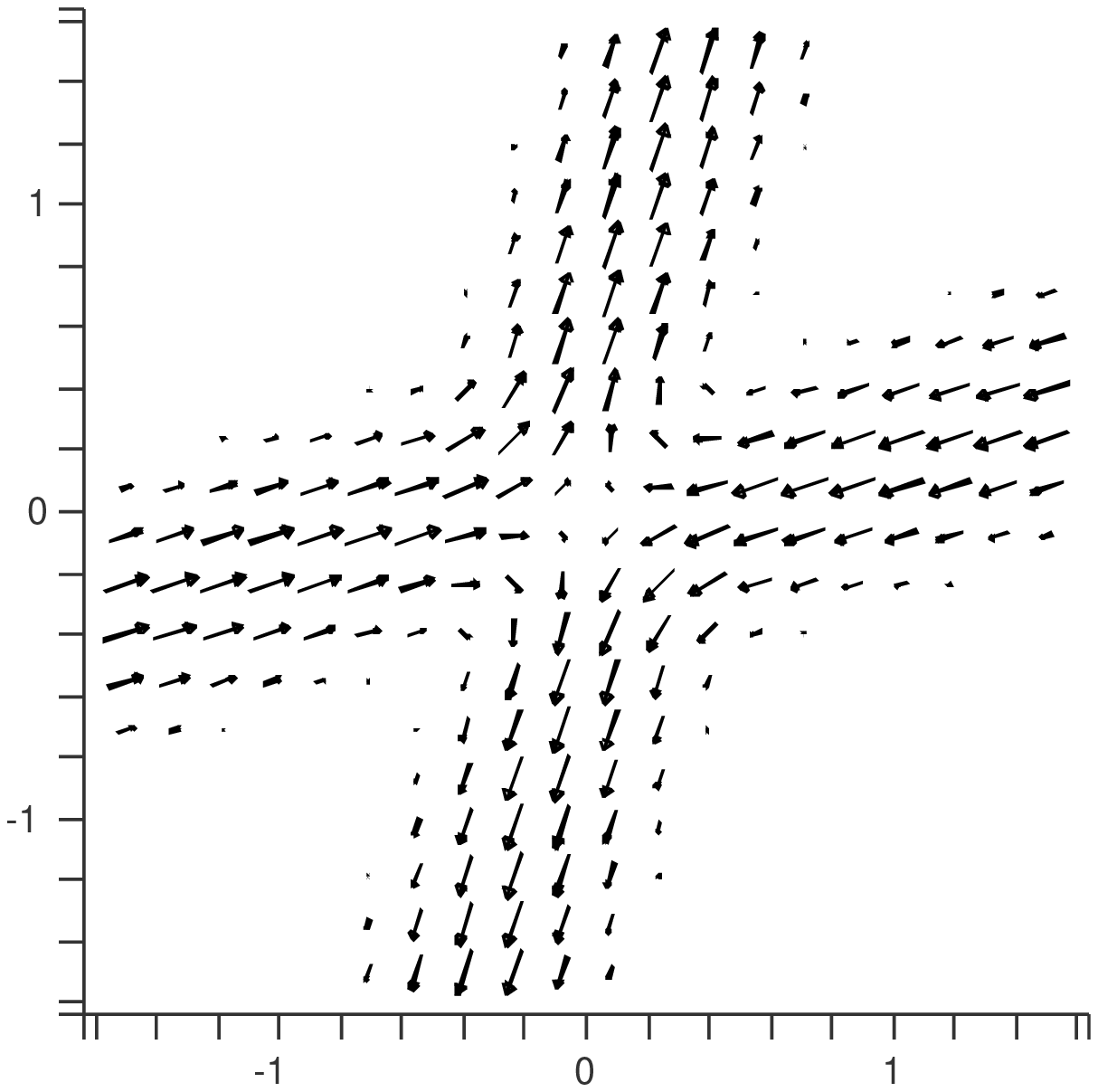}
  \caption{\label{fig:gyreSaddle}
           Gyre-saddle example. Standard configuration at $t=0$ (left) and skew configuration at $t=1/4$ (right).}
\end{figure}

A time-dependent variant of the linear saddle-type vector field
\begin{equation*}
\bu(\bx) =
\left(\begin{array}{l l l}
\label{eq:lin-saddle}
 \,\,\,\,\,\,\, a x \\
 - b y
\end{array}\right) \quad a > 0, \quad b > 0,
\end{equation*}
which is often used as a simple test case in vector field topology,
would be a straightforward approach for our analysis.
However, this field exhibits constant velocity gradient and because all trajectories separate 
at the same rate
in this vector field, it also exhibits constant FTLE. This is sometimes regarded as a drawback of Lagrangian methods such as FTLE, but due to the constant velocity gradient we regard it as a degenerate case in the Lagrangian view.
Therefore, we will be using a simple example derived from the so-called steady-type double-gyre flow.
This vector field is two-dimensional, divergence-free, spatially periodic, features a saddle point, and expresses FTLE ridges due to non-linearity \cite{rpbib:shadden2005tutorial}.
The derived example is made spatially non-periodic for simplicity, is still divergence-free,
% in the central region of diverging trajectories,
% (except for null sets at the boundaries),
 and offers a parameter for manipulating the angle at which the manifolds cross each other,
which will be varied
to obtain an exemplary time-dependent vector field.
We call our example gyre-saddle and define it inside the region 
% $D = [-1/2,1/2] \times [-1/2,1/2]$
$D = [-\frac{1}{2},\frac{1}{2}] \times [-\frac{1}{2},\frac{1}{2}]$
as
\begin{equation*}
\bu(\bx) =
\left(\begin{array}{l l l}
\label{eq:gyre-saddle-core}
 - \mathrm{sin}(\pi x) \mathrm{cos}(\pi y) + a \mathrm{sin}(\pi y) \mathrm{cos}(\pi x) \\
 \;\;\;\, \mathrm{sin}(\pi y) \mathrm{cos}(\pi x) - a \mathrm{sin}(\pi x) \mathrm{cos}(\pi y)
\end{array}\right)
\end{equation*}
and outside the region $D$ using
% $k=\mathrm{sgn}(y > |x|)$ and  $l=\mathrm{sgn}(x > |y|)$ as
 $k=(y \geq |x|) - (y \leq - |x|)$ and  $l=(x > |y|) - (x < - |y|)$,
with $(\cdot)$ being $1$ if true and $0$ if false,
 as
\begin{equation*}\label{eq:gyre-saddle}
\bu(\bx) =
\left(\begin{array}{c}
k a \mathrm{c}(\pi x - a \pi (y - \frac{k}{2})) - l \mathrm{c}(\pi y - a \pi (x - \frac{l}{2}))\\
k  \mathrm{c}(\pi x - a \pi (y - \frac{k}{2})) - l a \mathrm{c}(\pi y - a \pi (x - \frac{l}{2})) \end{array}\right)
\end{equation*}
with the clamped 
cosine
\begin{equation*}\label{eq:cos-clamp}
\mathrm{c}(x) = \left \{
\begin{array}{rl}
0 & \textrm{if }\,\,\, x < - \pi / 2 \\
\mathrm{cos}(x) & \textrm{if }\,\,\, - \pi / 2 \leq x \leq \pi / 2\\
0 & \textrm{if }\,\,\, x > \pi / 2 \\
%\mathrm{cos}(x) & \textrm{otherwise} 
\end{array} \right.
,
\end{equation*}
%TODO: clamping cos() at $-\pi/2$ and $\pi/2$ to $0$ is only $C^0$ continuous. Is this a problem?\\
and the parameter $a$ which is used to control the skew of the vector field. The standard configuration is 
$a = 0$. We will vary $a$ sinusoidally with time: $a(t) = \sin(2 \pi t) / 3$ (Figure~\ref{fig:gyreSaddle}).

\begin{figure}[tb]
  \centering
\begin{overpic}[width=1\linewidth]{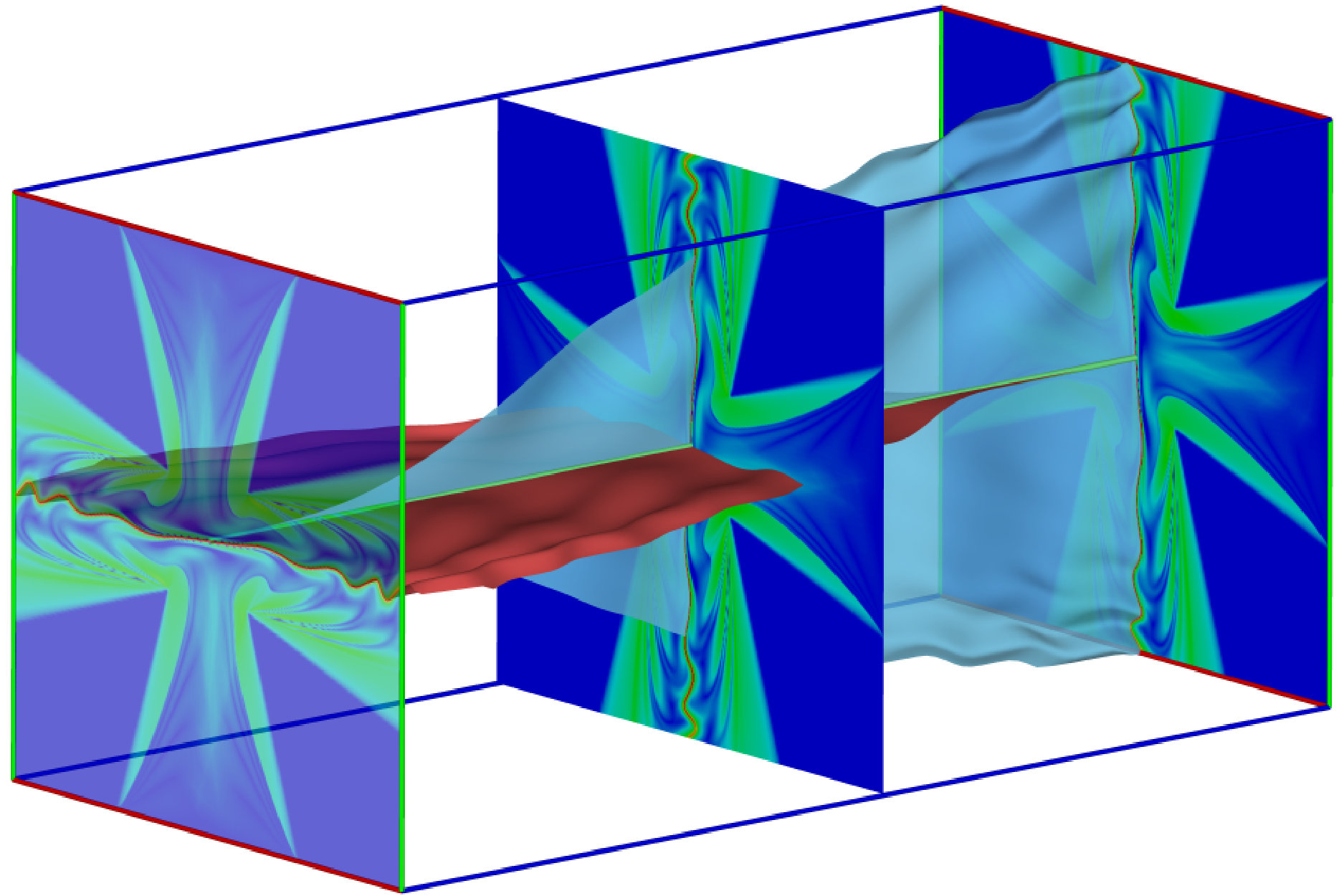}
\put(0,53) {
  \begin{tikzpicture}
  \coordinate [label=left:$t$] (A) at (0,0);
  \coordinate (B) at (0.5,0.1);
  \draw [->, thick] (A) -- (B);
  \end{tikzpicture}
}
\put(25, -3.7) { $t_0=4$ }
\put(58, 2.2) { $t_0=6.6$ }
\put(87.8, 8) { $t_0=9$ }
\put(0, 0) { \scriptsize{FTLE} }
\put(1,-3.4) {\includegraphics[width=0.12\linewidth]{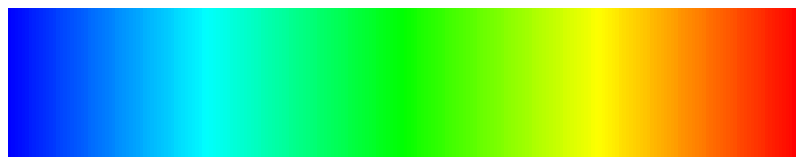}}
\put(0, -6.4) { \scriptsize{$0$}}
\put(10.7, -6.4) { \scriptsize{$1$}}
\end{overpic}\\
\vspace{0.075\linewidth}
\begin{overpic}[width=1\linewidth]{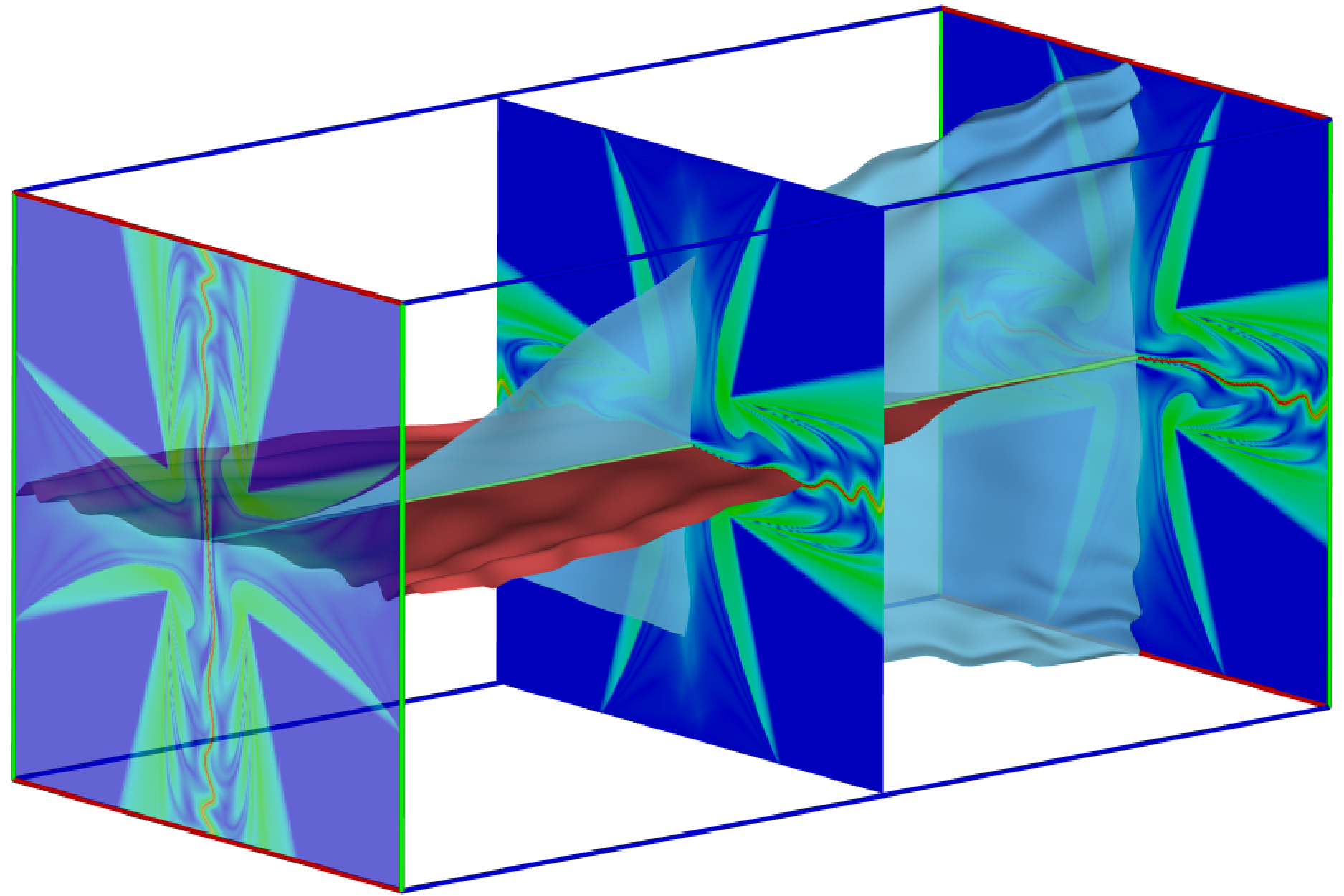}
\put(0,53) {
  \begin{tikzpicture}
  \coordinate [label=left:$t$] (A) at (0,0);
  \coordinate (B) at (0.5,0.1);
  \draw [->, thick] (A) -- (B);
  \end{tikzpicture}
}
\put(25, -3.7) { $t_0=4$ }
\put(58, 2.2) { $t_0=6.6$ }
\put(87.8, 8) { $t_0=9$ }
\end{overpic}
\vspace{0.01\linewidth}
  \caption{\label{fig:gyreSaddle-skewing1}
           Space-time visualization of skewing gyre-saddle example. Top: standard visualization. FTLE at time $t_0 = 4$ with advection time $T=5$, at time $t_0 = 6.6$ with advection time $T=-5$, and at time $t_0 = 9$ with advection time $T=-5$. Space-time streak manifolds (attracting: light blue, repelling: red) seeded along uniformly hyperbolic trajectory (green tube) are consistent with respective LCS. Bottom: visualization for verification with reversed FTLE advection times. The repelling LCS
% (FTLE images) 
should preferably intersect the attracting space-time streak manifold at the uniformly hyperbolic trajectory. This is the case over the complete time interval in this example. }
\end{figure}

Figure~\ref{fig:gyreSaddle-skewing1} shows a visualization of the skewing saddle example.
A seed for a uniformly hyperbolic trajectory inside the visualized time interval was found and space-time streak manifolds are visualized together with backward-time FTLE fields for verification of the attracting streak manifold by the corresponding attracting LCS. 
Because the vector field exhibits a temporal period of $\tau=1$, the intermediate FTLE was chosen at $t_0=6.6$ for reliable validation.
It can be seen that it is consistent. For a verification of the repelling streak manifold, forward-time FTLE fields are
% visualized in Figure~\ref{fig:gyreSaddle-skewing2}. 
also visualized.
Again, the LCS and the respective streak manifold are consistent. 
The visualization in Figure~\ref{fig:gyreSaddle-skewing1}~(bottom)
also serves as a validation of the uniformly hyperbolic trajectory: it should stay on the intersection curve of the invariant manifolds and therefore the attracting streak manifold should intersect the repelling LCS at the uniformly hyperbolic trajectory. It can be seen that this is the case for the visualized time interval in this example.

%- - - - - - - - - - - - - - - - - - - - - - - - - - - - - - - - - - - - -
\subsection{Oscillating gyre-saddle example}
\label{sec:gyre-saddle-oscillating}

\begin{figure}[tb]
  \centering
\begin{overpic}[width=1\linewidth]{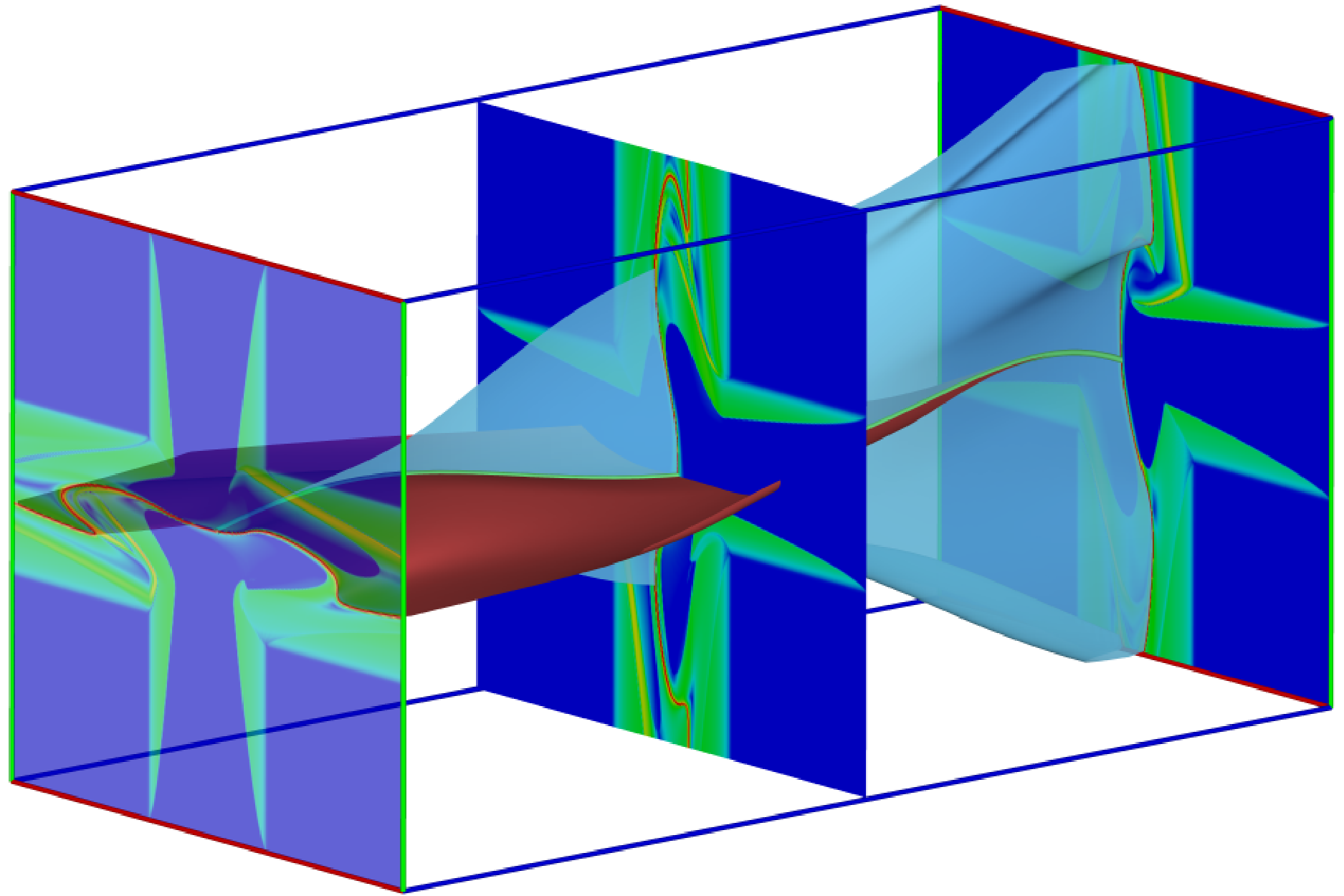}
\put(0,53) {
  \begin{tikzpicture}
  \coordinate [label=left:$t$] (A) at (0,0);
  \coordinate (B) at (0.5,0.1);
  \draw [->, thick] (A) -- (B);
  \end{tikzpicture}
}
\put(25, -3.7) { $t_0=4$ }
\put(58, 2.2) { $t_0=6.5$ }
\put(87.8, 8) { $t_0=9$ }
\put(0, 0) { \scriptsize{FTLE} }
\put(1,-3.4) {\includegraphics[width=0.12\linewidth]{figures/gyreSaddle_skewing_2009-10-05_colorLegend}}
\put(0, -6.4) { \scriptsize{$0$}}
\put(8.2, -6.4) { \scriptsize{$0.9$}}
\put(86.5,52) {
  \begin{tikzpicture}
  \coordinate (A) at (0,0);
  \coordinate (B) at (.31622,0);
  \draw [->, white, very thick] (B) -- (A);
  \end{tikzpicture}
}
\end{overpic}\\
\vspace{0.075\linewidth}
\begin{overpic}[width=1\linewidth]{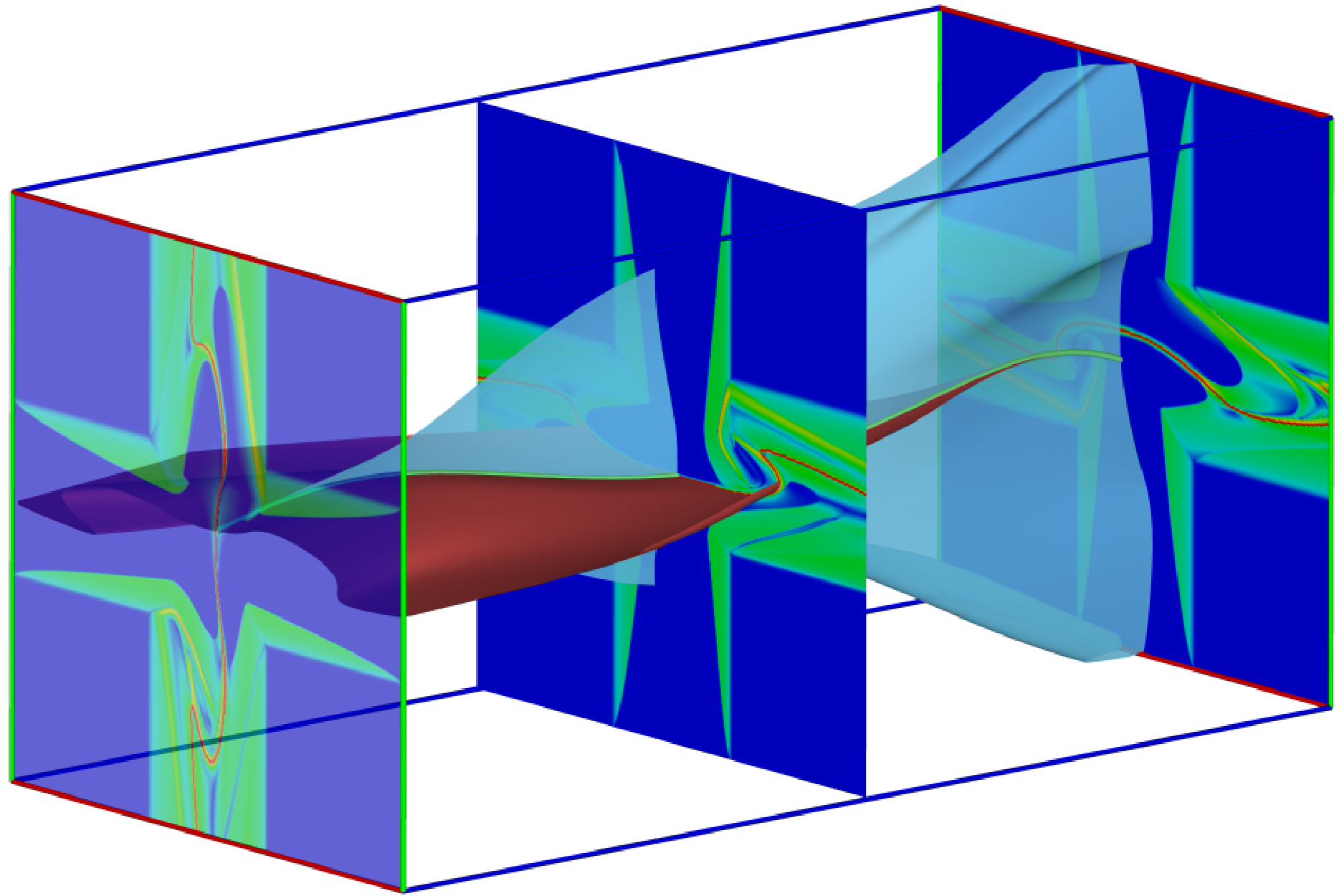}
\put(0,53) {
  \begin{tikzpicture}
  \coordinate [label=left:$t$] (A) at (0,0);
  \coordinate (B) at (0.5,0.1);
  \draw [->, thick] (A) -- (B);
  \end{tikzpicture}
}
\put(25, -3.7) { $t_0=4$ }
\put(58, 2.2) { $t_0=6.5$ }
\put(87.8, 8) { $t_0=9$ }
\put(83.2,36.8) {
  \begin{tikzpicture}
  \coordinate (A) at (0,.2236);
  \coordinate (B) at (0.2236,0);
  \draw [->, white, very thick] (B) -- (A);
  \end{tikzpicture}
}
\end{overpic}
\vspace{0.01\linewidth}
  \caption{\label{fig:gyreSaddle-sineTransl3}
           Space-time visualization of oscillating gyre-saddle example. Top:
% standard visualization. 
FTLE at time $t_0 = 4$ with advection time $T=5$, at $t_0 = 6.5$ with $T=-5$, and at $t_0 = 9$ with $T=-5$. Space-time streak manifolds (attracting: light blue, repelling: red) seeded along uniformly hyperbolic trajectory (green tube) are consistent with respective LCS. Additional shear-induced LCS (arrow) are not captured. Bottom: reversed FTLE advection times for verification. The repelling LCS should preferably intersect the attracting streak manifold at the uniformly hyperbolic trajectory.  This is still the case at $t=6.5$ but not anymore at $t=9$ (arrow). Please see Section~\ref{sec:gyre-saddle-oscillating} for a discussion of this issue. }
\end{figure}

Here, we apply our method to an other variant of the gyre-saddle dataset from Section~\ref{sec:crit-non-moving}. Instead of a time-dependent skew, the example is at standard configuration ($a=0$), and is made time-dependent by a harmonic oscillation of the saddle diagonally between the locations $(0.25,0.25)$ and $(-0.25,-0.25)$ with a period of $\tau = 4$. 
%Figures~\ref{fig:gyreSaddle-sineTransl3} and \ref{fig:gyreSaddle-sineTransl4} show 
Figure~\ref{fig:gyreSaddle-sineTransl3} shows
the space-time visualization. The attracting LCS in the FTLE
% (Figure~\ref{fig:gyreSaddle-sineTransl3})
% at time $t_0=9$ is
is consistent with the attracting space-time streak manifold over the entire time interval but the additional
% two 
boundary-shear induced attracting LCS
% at the top right and mid-bottom left 
are not captured because they are not related to hyperbolic mechanisms.

For a verification of the repelling streak manifold, forward-time FTLE fields are also visualized
% (Figure~\ref{fig:gyreSaddle-sineTransl4}). 
(Figure~\ref{fig:gyreSaddle-sineTransl3} (bottom)). 
Again, the LCS and the streak manifold are consistent. However, the trajectory does not stay over the complete visualized time interval on the space-time intersection curve of the attracting and repelling LCS (compare visualization at time $t=9$). Nevertheless, it starts to deviate only at the end of the interval.
We tried to address this problem by extracting the seed at very high precision from ridges of long-time FTLE (advection time $|T|=7.9$) and also hyperbolicity time, but this did not lead to better results. Also 
%using very small integration steps 
performing very precise integration
for the computation of FTLE and hyperbolicity time as well as for the uniformly hyperbolic trajectory did not reduce the problem. Since the problem seems to be hard to isolate, we find it an interesting topic for future research. Still, it has to be noted that the visualization typically does not suffer because the streak manifolds get quickly attracted by the
% invariant manifold in the respective direction of time.
respective invariant manifold.

%- - - - - - - - - - - - - - - - - - - - - - - - - - - - - - - - - - - - -
\subsection{Quad-gyre example}
\label{sec:quad-gyre}

\begin{figure}[htb]
  \centering
  \includegraphics[width=.58\linewidth]{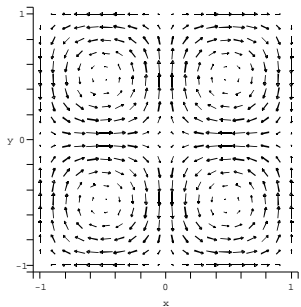}
  \caption{\label{fig:quadGyre}
           Quad-gyre example at $t=0$. }
\end{figure}

Here, we apply the method to the so-called time-dependent double-gyre example due to Shadden~\cite{rpbib:shadden2005tutorial}. It is temporally and spatially periodic and we use a larger range of the field, resulting in four gyres, hence we call this example the quad-gyre. Using
\begin{align*}
f(x,t) & = a(t) x^2 + b(t) x,\\
a(t) & = \epsilon \sin(\omega t),\\
b(t) & = 1 - 2 \epsilon \sin(\omega t),
\end{align*}
it is defined as follows:
\begin{equation*}
\bu(\bx) =
\left(\begin{array}{l l l}
\label{eq:quad-gyre-2}
- \pi A \sin(\pi f(x)) \cos(\pi y)\\
\pi A \cos(\pi f(x)) \sin(\pi y) \frac{d f}{ d x} 
\end{array}\right)
.
\end{equation*}
We use the configuration $\epsilon=0.25$, $\omega = 2 \pi / 10$, and $A=0.1$.
Figure~\ref{fig:quadGyre} shows a plot at $t=0$. The saddle-type critical points at $x=0$ oscillate horizontally while those at $x=-1$ and $x=1$ are stationary.

\begin{figure*}[tb]
  \centering
\begin{overpic}[width=1\linewidth]{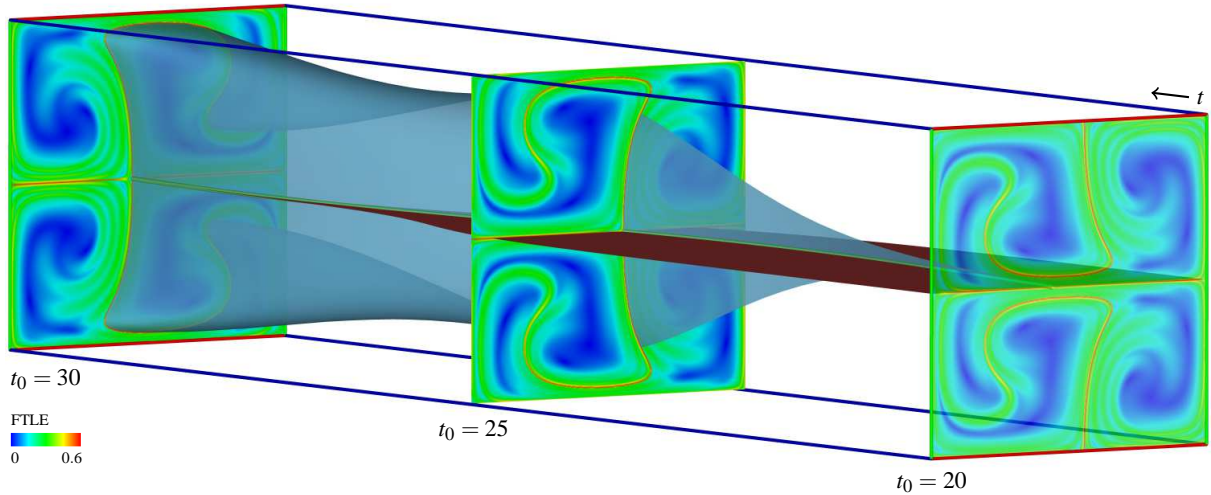}
\put(94.5,30) {
  \begin{tikzpicture}
  \coordinate [label=right:$t$] (A) at (0,0);
  \coordinate (B) at (-0.5,0.06);
  \draw [->, thick] (A) -- (B);
  \end{tikzpicture}
}
\put(73.5, -1.2) { $t_0=20$ }
\put(35.5, 3.1) { $t_0=25$ }
\put(0, 7.3) { $t_0=30$ }
\put(0, 4) { \scriptsize{FTLE} }
\put(0.4,2.2) {\includegraphics[width=0.06\linewidth]{figures/gyreSaddle_skewing_2009-10-05_colorLegend}}
\put(0, 0.8) { \scriptsize{$0$}}
\put(4.2, 0.8) { \scriptsize{$0.6$}}
\end{overpic}
\vspace{0.0018\linewidth}
  \caption{\label{fig:quad-gyre-1}
           Space-time visualization of time-dependent quad-gyre example. FTLE at time $t_0 = 20$ with advection time $T=10$, at $t_0 = 25$ with advection time $T=-10$, and at $t_0 = 30$ with advection time $T=-10$. Space-time streak manifolds (attracting: light blue, repelling: red) seeded along uniformly hyperbolic trajectory (green tube) are consistent with respective LCS over the entire time interval.
}
\end{figure*}

Figure~\ref{fig:quad-gyre-1} shows the visualization using our method. A seed for a uniformly hyperbolic trajectory inside the visualized time interval was found and space-time streak manifolds are visualized together with backward-time FTLE fields for verification of the attracting streak manifold by the corresponding attracting LCS. It can be seen that it is consistent. Because the $y$-symmetry of the example is not varying over time, we do not show additional forward-time FTLE for verification of the repelling streak manifold. It is consistent with the $y=0$ plane and hence consistent with the corresponding repelling LCS.
The uniformly hyperbolic trajectory also stays on that plane and hence follows the intersection curve of the invariant manifolds.

%- - - - - - - - - - - - - - - - - - - - - - - - - - - - - - - - - - - - -
\subsection{Buoyancy example}
\label{sec:buoyancy}

\begin{figure}[htb]
  \centering
  \includegraphics[width=.48\linewidth]{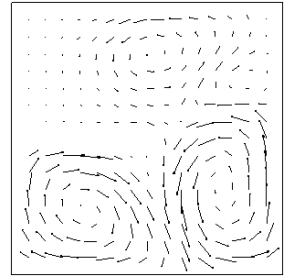}
  \caption{\label{fig:buoyancy-hog}
           Buoyancy CFD example at $t=2$. }
\end{figure}

This example is a 2D time-dependent CFD simulation of buoyant air. The bottom wall is heated whereas the upper one is cooled, while the left and right walls are neutral. There is a horizontal barrier extending from the mid of the left wall to the center (Figure~\ref{fig:buoyancy-hog}). Convective flows are known to exhibit very complicated LCS that drive the mixing in the flow, i.e.~by so-called thinning and folding.

Unfortunately we were not able to extract seeds for uniformly hyperbolic trajectories in significant regions and for significant time intervals, as well as in another CFD example. It has to be subject to further investigation if this is due to numerics, or if these are cases where the method is not applicable; Haller stated \cite{Haller2000invariantManif} that his method is only applicable if the deformation rate of the LCS is lower than typical particle speeds. Therefore, we follow our ``weak'' hyperbolicity approach discussed in Section~\ref{sec:hyp-time-and-ridges}, meaning that we generate streak manifolds as long as the seeding trajectory stays in a hyperbolic region.
% Another reason for this approach is that 
We obtain the seeds from moderate-time ($|T|=0.5$ seconds, same as the visualization time interval) FTLE fields because
long-time FTLE and hyperbolicity time exhibit massively folded ridges (see Figure~\ref{fig:hyp-time-quantization}) that would lead to a very high number of seeds and hence insignificant visualizations.
% Therefore we extract the seeds from low-time ($0.5$ seconds, same as the visualization time interval) FTLE ridges.

\begin{figure*}[tb]
  \centering
\begin{overpic}[width=1\linewidth]{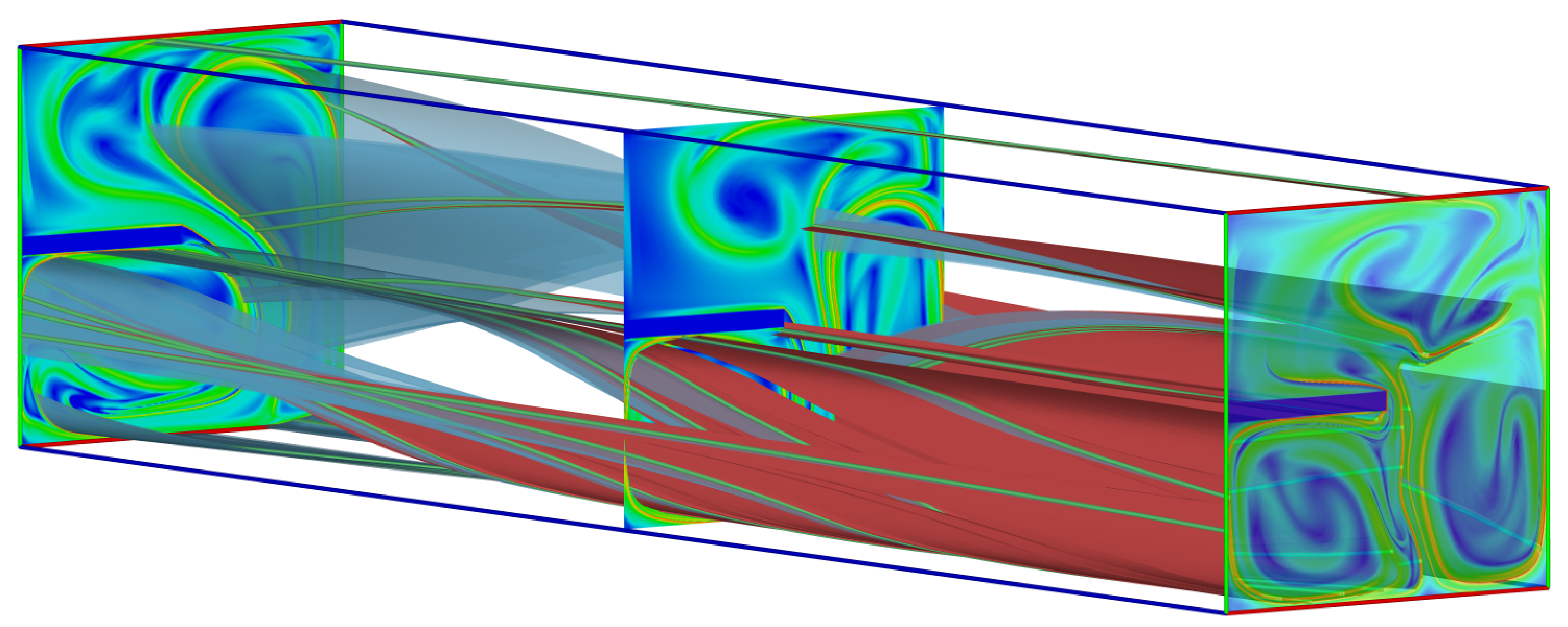}
\put(93.5,28.5) {
  \begin{tikzpicture}
  \coordinate [label=right:$t$] (A) at (0,0);
  \coordinate (B) at (-0.5,0.06875);
  \draw [->, thick] (A) -- (B);
  \end{tikzpicture}
}
\put(75.5, -1.4) { $t_0=2$ }
\put(36.7, 3.3) { $t_0=2.25$ }
\put(0.5, 8.4) { $t_0=2.5$ }
\put(0.6, 4) { \scriptsize{FTLE} }
\put(1,2.2) {\includegraphics[width=0.06\linewidth]{figures/gyreSaddle_skewing_2009-10-05_colorLegend}}
\put(0.6, 0.8) { \scriptsize{$0$}}
\put(4.8, 0.8) { \scriptsize{$0.6$}}
\end{overpic}\\
\vspace{0.03\linewidth}
\begin{overpic}[width=1\linewidth]{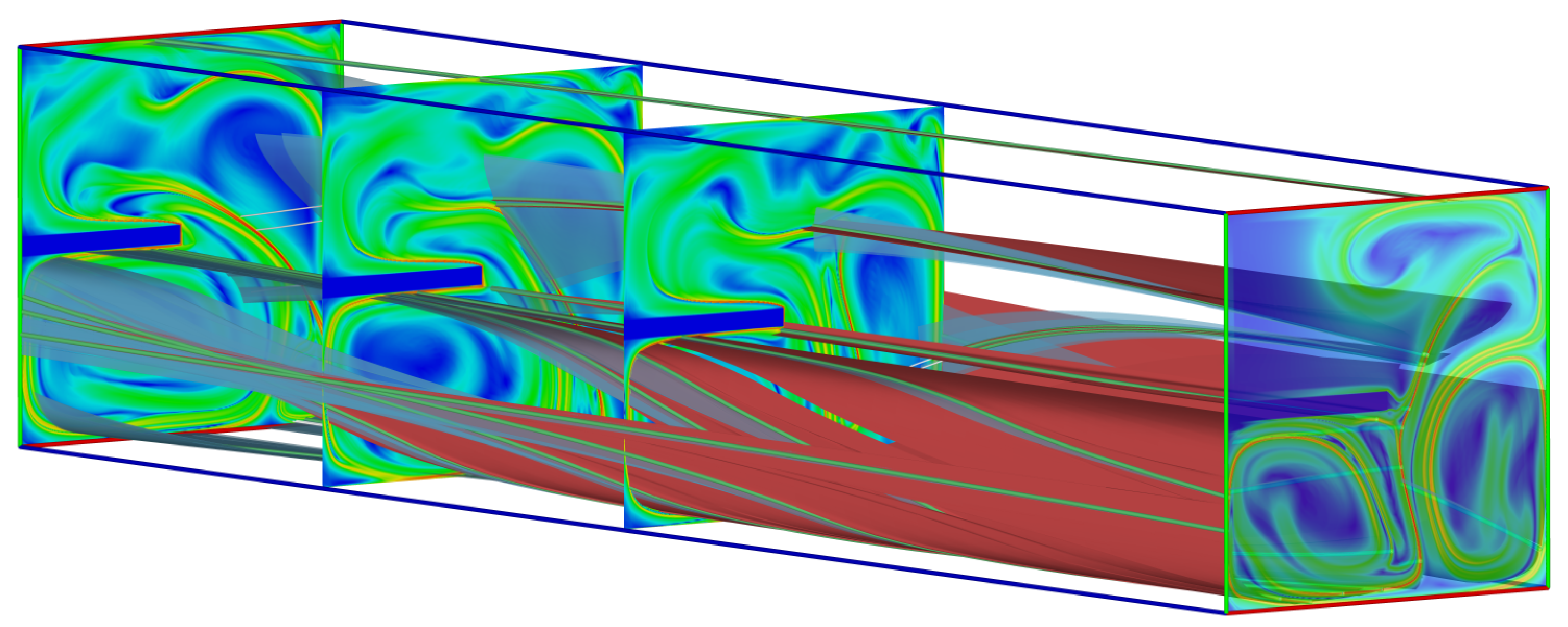}
\put(93.5,28.5) {
  \begin{tikzpicture}
  \coordinate [label=right:$t$] (A) at (0,0);
  \coordinate (B) at (-0.5,0.06875);
  \draw [->, thick] (A) -- (B);
  \end{tikzpicture}
}
\put(75.5, -1.4) { $t_0=2$ }
\put(36.7, 3.3) { $t_0=2.25$ }
\put(17.4, 5.8) { $t_0=2.375$ }
\put(0.5, 8.4) { $t_0=2.5$ }
\put(29,27.2) {
  \begin{tikzpicture}
  \coordinate (A) at (0,.2236);
  \coordinate (B) at (.2236,0);
  \draw [->, white, very thick] (A) -- (B);
  \end{tikzpicture}
}
\put(15.5,26.7) {
  \begin{tikzpicture}
  \coordinate (A) at (0,.2236);
  \coordinate (B) at (.2236,0);
  \draw [->, black, very thick] (A) -- (B);
  \end{tikzpicture}
}
\end{overpic}
\vspace{0.001\linewidth}
  \caption{\label{fig:buoyancy-1}
           Space-time visualizations of buoyancy example. Top: FTLE at time $t_0 = 2$ with advection time $T=0.5$, at $t_0 = 2.25$ with $T=-0.5$, and at $t_0 = 2.50$ with $T=-0.5$. Space-time streak manifolds (attracting: light blue, repelling: red) seeded along weakly hyperbolic trajectories (green tubes) are consistent with respective LCS, although some would need longer advection times to cover the full extent of the LCS. Bottom:
% FTLE at time $t_0 = 2$ and advection time $T=-0.5$ (right), at time $t_0 = 2.25$ and advection time $T=0.5$ (middle), $t_0 = 2.375$ and advection time $T=0.5$ (middle left), and $t_0 = 2.50$ and advection time $T=0.5$ (left).
FTLE with reversed advection times and additionally at $t_0 = 2.375$ with advection time $T=0.5$.
 Space-time streak manifolds are stopped if trajectory gets non-hyperbolic (visualized by white tubes, black arrow). Repelling LCS
% (FTLE images) 
should preferably intersect attracting streak manifolds at hyperbolic trajectories. This is still mostly the case at $t=2.25$ but not necessarily at later times (white arrow),
%.
%, because the trajectories are not uniformly hyperbolic, limited accuracy of the seeds, and due to integration error.
%This may be due to the fact that the trajectories are not uniformly hyperbolic, but this problem was also observed with uniformly hyperbolic trajectories (Figure~\ref{fig:gyreSaddle-sineTransl4}), 
see Section~\ref{sec:gyre-saddle-oscillating} for a discussion.
 }
\end{figure*}

Figure~\ref{fig:buoyancy-1} shows two visualizations of the data. The space-time streak manifolds are visualized together with backward-time FTLE fields for verification of the attracting streak manifolds by the corresponding attracting LCS. It can be seen that they are consistent. For a verification of the repelling streak manifolds, forward-time FTLE fields are also visualized. Again, the LCS and the streak manifolds are consistent. However, as in the example from Section~\ref{sec:gyre-saddle-oscillating} the seeding trajectories do not stay on the space-time intersection curves of attracting and repelling LCS over the complete visualized time interval. Nevertheless, they keep close in the first half of the visualized time interval. As the trajectory from Section~\ref{sec:gyre-saddle-oscillating} was determined to be uniformly hyperbolic, in contrast to our weakly hyperbolic trajectories used here,
but also exhibits deviation from the intersection curve,
 we see 
this fact as a motivation for our approach.
Additionally, the resulting streak manifolds still got attracted to the respective invariant manifold in all our experiments.

%-------------------------------------------------------------------------
\section{Conclusion}
\label{sec:conclusion}

We presented an approach to a time-dependent vector field topology. The concept is inspired by Lagrangian coherent structures (LCS) present as ridges in the finite-time Lyapunov exponent (FTLE). We mainly build on the findings by Haller \cite{Haller2000invariantManif} about invariant manifolds of hyperbolic trajectories in time-dependent vector fields: his concept can be directly reinterpreted as a time-dependent vector field topology when the role of streamlines is substituted by generalized streak lines. In this sense one can say that vector field topology should have been formulated based on generalized streak lines, since they are identical to streamlines in stationary vector fields. However, we have to point out that our approach is not complete in the sense that we only found a time-dependent counterpart to critical points of type saddle, the hyperbolic trajectories in space-time, and their separatrices, the space-time streak manifolds. Therefore, only LCS regions that are related to hyperbolic mechanisms are represented, e.g.~LCS due to shear processes are not captured.
% Nevertheless, this can also be seen as an advantage: this way it classifies LCS.

Our approach is not restricted to Haller's formulation on the basis of ridges in hyperbolicity time, we
% also allow to 
propose to alternatively
base the concept on ridges in the FTLE. This allows for a 
robust and
scale-dependent approach by variation of the advection time used for FTLE computation and for significant visualizations that are consistent with LCS defined by ridges in the FTLE at the chosen advection time scale.

\noindent The presented approach exhibits the following advantages over the extraction of LCS from the FTLE:
\begin{itemize}
\item
Efficient computation of time series: whereas the FTLE needs to get recomputed for each frame of an animation, which is typically very time-consuming, the streak manifolds can be computed at arbitrary temporal and spatial resolution at comparably little computational cost.
\item
Ridge extraction is avoided: the resulting streaks are smooth and insusceptible to (numerical) noise.
% and numerical error.
\item
Hyperbolic trajectories give additional insight into the organization and dynamics of LCS.
\end{itemize}

\noindent But it also exhibits the following drawbacks compared to LCS by FTLE:
\begin{itemize}
\item
The approach relies on a sampled field 
for seeding the streak manifolds
at discrete time steps: the FTLE or hyperbolicity time. Short-lived LCS may therefore get missed, i.e.~hyperbolic trajectories that are short in space-time. This can lead to popping artifacts over time.
\item
The FTLE is not computed on the space-time streak manifolds and is therefore not available for a quantitative interpretation of the LCS regarding attraction or repulsion.
\item
Hyperbolic trajectories are hard to integrate: errors in the seeding position or during integration tend to grow exponentially.
\end{itemize}

We identified several topics for future research. Finding time-dependent counterparts to critical points of type node, focus, and center, as well as periodic orbits is probably the most evident but maybe also most demanding step. A 
robust and efficient method for extracting highly accurate ridge intersections in case of massively folded ridges is probably easier to achieve and therefore the next step we address. Finally, the approach could be extended to 3D vector fields.

%-------------------------------------------------------------------------
\section{Acknowledgments}
\label{sec:acknowledgements}

We would like to thank Ronald Peikert (ETH Zurich) for his support.

%-------------------------------------------------------------------------

\bibliographystyle{eg-alpha}
%\bibliography{timeVFT,rpbib}
\bibliography{timeVFT}

\end{document}